\documentclass[11pt,english]{article}

\usepackage[T1]{fontenc}
\usepackage[utf8]{inputenc}

\makeatletter

\usepackage{titling}

\usepackage[margin=.85in]{geometry}

\usepackage{booktabs}
\usepackage[english]{babel}
\usepackage{subcaption}

\usepackage{fixltx2e}
\usepackage{dcolumn} 
\usepackage[flushleft]{threeparttable}
\usepackage{threeparttablex}

\usepackage{babel}
\usepackage[authoryear,round,longnamesfirst]{natbib}
\setcitestyle{aysep={},citesep={;}} 

\usepackage{enumitem}
\setlist{nolistsep}

\usepackage{longtable}

\usepackage{nicefrac}

\usepackage[labelfont={sc},textfont=normalfont]{caption}

\usepackage{setspace}
\usepackage{indentfirst}

\usepackage{pdflscape}

\usepackage{rotating}

\usepackage{blindtext,lipsum}
\usepackage{microtype,enumitem}
\usepackage{setspace}\setdisplayskipstretch{}

\setlist{noitemsep}
\usepackage{bbm}

\usepackage[multiple]{footmisc}
\usepackage{url}

\usepackage[compact]{titlesec}

\usepackage[dvipsnames,svgnames,x11names,hyperref]{xcolor}

\definecolor{red}{rgb}{0.502,0,0}

\newcommand{\xfnm}[1][]{\ifx!#1!\else\unskip,\space#1\fi}

\usepackage{threeparttable}
\usepackage{pgfplots}
\usepackage{xcolor}
\pgfplotsset{compat=1.18}

\definecolor{barblue}{RGB}{66,133,244}
\definecolor{overallcolor}{RGB}{219,68,55}
\definecolor{gridcolor}{gray}{0.85}

\usepackage{float}
\usepackage{morefloats}
\usepackage{graphics}

\usepackage{url}
\usepackage{rotating}
\usepackage{multirow}
\usepackage[section]{placeins}
\usepackage{tabularx}
\usepackage{lscape}
\usepackage{paralist} 
\usepackage{verbatim} 
\usepackage{setspace}
\usepackage{csquotes}

\usepackage{multibib}
\newcites{app}{Appendix References}
\newcites{oapp}{Online Appendix References}
\newcites{smapp}{Supplementary Material References}

\usepackage{amsmath}
\usepackage{amssymb}
\usepackage{amsfonts}
\usepackage{amsthm}
\usepackage{array} 
\usepackage{bbm}
\usepackage{mathrsfs}
\usepackage{mathbbol}

\usepackage{pxfonts}

\usepackage{sectsty}
\allsectionsfont{\rmfamily\bfseries\upshape} 

\newcommand\namepaper{\textbf{Human-AI Collaboration in Radiology: \\The Case of Pulmonary Embolism}}

\usepackage{hyperref}

\hypersetup{
	bookmarks=true,
	bookmarksopen=false,
	colorlinks=true,
	urlcolor=red,
	linkcolor=red,
	citecolor=red,
	pdfpagelayout=SinglePage,
	pdfpagemode=UseNone
}

\usepackage[capitalise,nameinlink]{cleveref}

\makeatother

\begin{document}

\title{\namepaper\thanks{We are grateful to the AI developer's research team and the hospital system's data team for their extensive support. We thank Howie Forman and Melissa Davis for their helpful comments and guidance. Part of this research was conducted while Zentefis was visiting the Hoover Institution, Stanford University. This study was approved under Yale Institutional Review Board IRB-2000037424, MOD00071941.} }

\author{Paul Goldsmith-Pinkham\thanks{Yale School of Management; 165 Whitney Ave, New Haven, CT 06510 (email: paul.goldsmith-pinkham@yale.edu)} \and Chenhao Tan\thanks{The University of Chicago Department of Computer Science and Data Science; 5730 S Ellis Ave, Chicago, IL 60637 (email: chenhao@uchicago.edu)}  \and Alexander K. Zentefis\thanks{Leavey School of Business, Santa Clara University; 500 El Camino Real, Santa Clara, CA 95035 (email: azentefis@scu.edu), Corresponding author.}\\[1cm]}


\begin{titlepage}

\clearpage
\maketitle
\thispagestyle{empty}

\begin{abstract}
    We study how radiologists use AI to diagnose pulmonary embolism (PE), tracking over 100,000 scans interpreted by nearly 400 radiologists during the staggered rollout of a real-world FDA-approved diagnostic platform in a hospital system. When AI flags PE, radiologists agree 84\% of the time; when AI predicts no PE, they agree 97\%. Disagreement evolves substantially: radiologists initially reject AI-positive PEs in 30\% of cases, dropping to 12\% by year two. Despite a 16\% increase in scan volume, diagnostic speed remains stable while per-radiologist monthly volumes nearly double, with no change in patient mortality---suggesting AI improves workflow without compromising outcomes. We document significant heterogeneity in AI collaboration: some radiologists reject AI-flagged PEs half the time while others accept nearly always; female radiologists are 6 percentage points less likely to override AI than male radiologists. Moderate AI engagement is associated with the highest agreement, whereas both low and high engagement show more disagreement. Follow-up imaging reveals that when radiologists override AI to diagnose PE, 54\% of subsequent scans show both agreeing on no PE within 30 days.
\end{abstract}

\begin{flushleft}
\textbf{\vfill{}
JEL classification:} C21, C88, D81, I10, J24, O33  \linebreak{}
\par\end{flushleft}
\end{titlepage}

\clearpage

\begin{spacing}{1.5}

\section{Introduction}
\label{sec:intro}

Recent years have seen widespread adoption of artificial intelligence (AI) tools in professional decision-making. Human resource managers now routinely consult AI systems for candidate screening \citep{dattner2019legal}, financial analysts use them to evaluate investment opportunities \citep{Roy2025}, and judges employ them to assess recidivism risk in bail decisions \citep{angelova2025algorithmic}. While studies have documented the remarkable accuracy and technical capabilities of these AI systems, the real-world dynamics of human-AI interaction and the resulting impact on professional decisions are not well understood. As AI collaboration becomes increasingly prevalent in professional work, more research is warranted on its influence on human decision-making.

In this paper, we attempt to make progress in this area by examining how radiologists collaborate with AI when diagnosing pulmonary embolism (PE). A PE is a blood clot that blocks arteries in the lungs \citep{tapson2005acute}. It is the third leading cause of cardiovascular death in the U.S., after heart attack and stroke \citep{duffett2020pulmonary}, killing an estimated 60-100,000 people in the U.S. per year \citep{freund2022acute}. We study suspected PE as opposed to incidental PE, which is detected inadvertently on imaging ordered for other reasons. When PE is suspected, computed tomographic pulmonary angiography (CTPA) is the clinical standard of care \citep{schoepf2004spiral,weiss2006ct,anderson2007computed,di2016deep}. Deep learning AI tools designed to analyze CTPA images have shown very high accuracy at detecting PEs \citep{soffer2021deep,weikert2020automated,huhtanen2022automated}, and they increasingly serve as decision aids to radiologists \citep{ebrahimian2022predictive,cheikh2022artificial,rothenberg2023prospective}.

We study the deployment of an FDA-approved AI assistance tool at a large academic health system. The system was rolled out in a staggered fashion across eight care sites between August 2019 and July 2022. We track 117,063 CTPA scans interpreted by 389 signing radiologists. We link these scans to AI predictions, radiologist diagnoses extracted from clinical reports, radiologist characteristics from national provider databases, and detailed engagement metrics showing how radiologists interacted with the AI system. This setting provides a unique opportunity to examine human-AI collaboration in actual clinical practice rather than experimental settings.

We document five main findings. First, radiologists show asymmetric agreement with AI predictions. When the AI predicts no PE, radiologists agree in 97\% of cases. When the AI flags PE, agreement drops to 84\%. This 13 percentage point gap reveals that radiologists more readily accept AI-negative assessments than AI-positive findings. 

Second, disagreement patterns evolve significantly after AI deployment. In the first year, radiologists reject AI-detected PEs in 30\% of cases. This drops sharply to 12\% by year two, then stabilizes. The reduction could reflect learning, as radiologists observe whether AI-flagged cases actually had PE, or automation bias, as radiologists gradually defer more to the system. Disagreement when AI predicts no PE remains consistently low at 2-3\% throughout. 

Third, despite a growing workload over this time period, diagnostic efficiency remains stable. The average time from order to diagnosis holds steady at 3.6 hours per scan even as total scan volume increased 16\%. Per-radiologist monthly volumes nearly double from 5.3 to 10.4 scans. Meanwhile, 30-day, 90-day, and 1-year patient mortality rates are roughly unchanged after the AI rollout. The stable per-scan reading time alongside higher throughput suggests productivity gains operate through optimizing workflow---faster case triage, reduced cognitive load from AI serving as a concurrent second reader, or more efficient handling of negative cases---rather than degraded decision-making that compromises patient health. Reading time for PE-positive cases increases from 3.2 to 3.6 hours, consistent with longer review of AI-flagged findings rather than rushed diagnosis.

Fourth, large heterogeneity persists across radiologists even after five years of AI use. Some radiologists reject AI-flagged PEs in half of cases. Others accept them in nearly all cases. The cross-sectional distribution of disagreement rates shifts over time but substantial variation remains. Female radiologists are 6 percentage points less likely to override AI-detected PEs than male radiologists. Experience shows no clear relationship with agreement patterns. Engagement with the AI system reveals a non-monotonic pattern: radiologists with moderate engagement (hovering over 26\% of AI alerts) show highest agreement at 91\% when AI flags PE. Both low-engagement radiologists (8\% hover rate) and high-engagement radiologists (45\% hover rate) show similar lower agreement at 81\%. This suggests two modes of disagreement: passive bypass at low engagement and active vetting at high engagement.

Fifth, follow-up imaging provides insight into disagreement resolution. Among 35,896 patients receiving initial CTPA scans with AI, 1,375 (3.8\%) return for follow-up imaging within 30 days. When radiologists initially override the AI to diagnose PE, 54\% of follow-up scans show both agreeing on no PE. When radiologists initially reject AI positive flags, none of the follow-up scans maintain that disagreement pattern. The findings demonstrate that radiologist overrides of AI negative predictions are more persistent than their rejections of AI positive flags.

These findings matter for three reasons. First, they reveal how professionals incorporate algorithmic recommendations in high-stakes decisions. The asymmetric agreement patterns, time evolution, and persistent heterogeneity document that human-AI collaboration involves active judgment rather than passive acceptance or rejection. Second, the findings provide evidence on the mechanisms through which AI affects professional work. The sharp initial reduction in disagreement followed by stabilization could reflect either beneficial learning or concerning automation bias. The non-monotonic relationship between engagement and agreement suggests that how professionals interact with AI systems shapes their collaboration patterns. The stable diagnostic speed alongside doubled per-radiologist volumes (without significantly higher patient mortality rates) indicates AI may enhance productivity through workflow optimization instead of faster, jeopardized individual decisions. Third, understanding these dynamics has implications for AI deployment in healthcare. The large heterogeneity across radiologists and the persistence of disagreement patterns even after years of use indicate that simply providing AI tools does not guarantee consistent utilization or collaboration patterns.

Overall, we find that radiologists and AI disagree in predictable but heterogeneous ways. Disagreement declines sharply after deployment but substantial variation persists across radiologists. Female radiologists show higher agreement with AI-detected PEs. Moderate engagement associates with highest concordance. Diagnostic efficiency remains stable despite increased workload. Follow-up imaging reveals asymmetric evolutions in diagnoses. Our planned future work will exploit the staggered rollout and quasi-random assignment of patients to radiologists to estimate the causal effects of AI assistance on radiologist diagnostic skill, preferences for balancing false positives and false negatives, and patient health outcomes including mortality, complications, PE treatment, and healthcare utilization.

\paragraph*{Related Literature.}

This paper connects to several research areas: AI in medical imaging, human-AI collaboration in professional work, and clinical decision support systems.

\textbf{AI in Radiology and Medical Imaging.} Deep learning has demonstrated high accuracy in detecting abnormalities across imaging modalities \citep{liu2019comparison,rajpurkar2023current}. For pulmonary embolism specifically, convolutional neural networks achieve a pooled sensitivity of 0.88 (the ability to correctly identify scans with PE) and a specificity of 0.86 (the ability to correctly identify scans without PE) on CTPA scans \citep{soffer2021deep}. Studies document AI systems correctly identifying PE with balanced accuracy \citep{weikert2020automated,grenier2023deep}. Despite technical success, evidence on real-world deployment effects remains limited. \citet{schmuelling2021deep} found no reduction in report communication times or patient turnaround nine months after AI implementation, highlighting the discrepancy between algorithmic performance and workflow benefits. We contribute by documenting how radiologists actually collaborate with AI systems in clinical practice over a five-year period, revealing patterns of agreement, disagreement, and evolution that algorithmic accuracy metrics alone cannot capture.

\textbf{Human-AI Collaboration.} Our work complements the experimental study by \citet{agarwal2023combining}, who conducted a lab-in-the-field experiment with 227 radiologists examining chest X-rays. While they find radiologists exhibit automation neglect and underweight AI signals in controlled settings, we examine human-AI collaboration in actual clinical deployment. Unlike their focus on probabilistic beliefs in experimental tasks, we investigate behavioral patterns in diagnoses affecting patient care. We document both the evolution of collaboration patterns over time and substantial heterogeneity across radiologists in how they incorporate AI recommendations.

Recent work emphasizes complementary skills between humans and AI \citep{patel2019human,shin2023impact}. \citet{leibig2022combining} show that combining radiologist and AI strengths in breast cancer screening improves diagnostic performance beyond either alone. Meta-analyses find AI assistance reduces reading time by 27\% while improving relative sensitivity by 12\% \citep{chen2024impact}. We extend this literature by examining actual collaboration patterns in deployed systems, documenting asymmetric agreement patterns and heterogeneity that experimental studies may not capture.

\textbf{Economics of Professional AI Adoption.} Economic research on AI in professional work examines whether AI substitutes for or complements human expertise \citep{brynjolfsson2017machine,acemoglu2018automation}. \citet{harris2024decision} shows that predictive AI helps expert technicians avoid unnecessary repairs in trucking. \citet{daugherty2018human} argue AI enables new division of labor where machines handle data processing while humans provide judgment and context. Our findings on radiologist override patterns and the non-monotonic relationship between engagement and agreement provide micro-level evidence of this complementarity. The persistence of disagreement even after years of use suggests that professional judgment continues to play an important role alongside algorithmic recommendations.

\textbf{Clinical Decision Support.} Medical AI functions as decision support rather than autonomous diagnosis \citep{who2021ethics,shortliffe2015clinical}. Studies document challenges in clinical integration including workflow disruption, an absence of trust, and interpretability concerns \citep{kelly2019key,geis2019ethics}. We contribute by measuring actual utilization patterns when radiologists have discretion to accept or override AI recommendations. Our findings help reveal how these integration challenges manifest in practice.

\textbf{Contribution.} We make four contributions. First, we provide detailed evidence on AI collaboration patterns in real clinical deployment rather than experimental settings. The large-scale data spanning five years of AI use, detailed engagement metrics, and follow-up imaging enable analysis of collaboration dynamics that controlled experiments cannot capture. Second, we document substantial heterogeneity in how radiologists incorporate AI recommendations. This heterogeneity persists even after years of use and relates to both observable characteristics like gender and behavioral patterns like system engagement. Third, we show that collaboration patterns evolve over time in ways consistent with either learning or automation bias. The sharp reduction in disagreement followed by stabilization provides evidence that professionals adapt their behavior as they gain experience with AI systems. Fourth, we show that disagreement patterns vary by direction. Radiologists accept AI-negative results more readily than AI-positive ones. Furthermore, their decisions to override a negative AI detection (and diagnose a PE) are more likely to persist in follow-up imaging than their decisions to reject an AI positive finding. These asymmetries suggest that human-AI collaboration involves more than simple acceptance or rejection of algorithmic recommendations.

\section{Setting and Data}
\label{sec:data}

\subsection{Pulmonary Embolism}
\label{sec:pe_background}

A pulmonary embolism (PE) is a blood clot that blocks arteries in the lungs. PE is the third leading cause of cardiovascular death in the United States, after heart attack and stroke \citep{duffett2020pulmonary}. The condition kills an estimated 60,000 to 100,000 Americans annually \citep{freund2022acute}. PE commonly forms when blood clots travel from veins in the legs or pelvis to the lungs through the bloodstream. Once lodged in the pulmonary arteries, these clots obstruct blood flow and reduce oxygen supply to lung tissue. Without treatment, PE mortality exceeds 30\% \citep{tapson2005acute}. With appropriate treatment, mortality drops below 10\% \citep{heit2015epidemiology}.

\Cref{fig:pe_anatomy} illustrates how blood clots form in leg veins, travel through the circulatory system, and obstruct the pulmonary arteries. The inset shows the embolus blocking blood flow and causing tissue damage. Clinical presentation varies from no symptoms to sudden death. Common symptoms include shortness of breath, chest pain, rapid heart rate, and coughing blood \citep{torbicki2008guidelines}. Risk factors include surgery, prolonged immobility, cancer, pregnancy, and inherited clotting disorders \citep{anderson2003risk}.

When clinicians suspect PE based on presenting symptoms, risk factors, and their clinical assessment, they order computed tomographic pulmonary angiography (CTPA). CTPA is the diagnostic standard for suspected PE \citep{schoepf2004spiral,weiss2006ct,anderson2007computed}. The scan injects contrast dye that makes blood vessels appear bright white on images. For a single CTPA, radiologists examine hundreds of cross-sectional images looking for filling defects---dark areas inside bright vessels where clots block blood flow. 

PE can also be detected incidentally on imaging ordered for other reasons, such as cancer staging, cardiac evaluation, or aortic imaging. Incidental PE accounts for approximately 1-3\% of all chest CT scans and represents 15-30\% of all PE diagnoses \citep{dentali2010prevalence,kearon2003natural,gladish2006incidental}. Our analysis focuses exclusively on suspected PE cases where CTPA was ordered with PE as the primary diagnostic target. This restriction reflects data feasibility. Identifying all incidental PE would require examining every chest CT scan regardless of indication---a substantially larger and more heterogeneous sample. The distinction between suspected and incidental PE is clinically important: patients with suspected PE typically present with acute symptoms and receive anticoagulation (blood thinning) treatment, whereas incidental PE patients are often asymptomatic and treatment decisions are more complex \citep{o2015treat}. All our results should be interpreted as pertaining to AI-assisted diagnosis of suspected PE and may not generalize to incidental PE detection.

\Cref{fig:ctpa_examples} shows two anonymized CTPA images from patients with PE. Panel A shows the upper chest where the main pulmonary artery splits. The grey area inside the right pulmonary artery (left side of image) is a blood clot blocking flow. The vessel should appear uniformly white from the contrast dye. Panel B shows the heart chambers. The right ventricle measures 56.4 mm compared to 41.0 mm for the left ventricle. Normally the left ventricle is larger. The enlarged right ventricle indicates the heart is working harder to pump blood through the blocked vessels. Together, Panel A reveals a clot's location while Panel B reveals the resulting heart strain.

\subsection{AI-Assisted Diagnosis}
\label{sec:ai_system}

Deep learning models achieve high accuracy in detecting PE on CTPA scans. Meta-analyses report pooled sensitivity of 0.88 (the ability to correctly identify scans with PE) and specificity of 0.86 (the ability to correctly identify scans without PE) \citep{soffer2021deep}. Several AI platforms have received FDA clearance for clinical use as decision support tools \citep{cheikh2022artificial}. These systems analyze CTPA images and flag suspected PE cases for radiologist review.

We study the deployment of an FDA-cleared AI platform at a large academic health system. The platform uses a convolutional neural network trained on over 15,000 CTPA scans with expert annotations. The algorithm processes each CTPA scan as it enters the radiology queue and generates a binary prediction: PE detected or no PE detected.

Radiologists work through a queue of scans that require interpretation. When the AI detects PE, it moves the scan to the top of this queue and displays a bright orange alert within the reading software. The interface shows a preview of the key image slice where the AI detected the finding, overlaid with a heat map highlighting the suspicious region. Radiologists can access the complete scan directly from the alert. When the AI does not detect PE, scans remain in their original queue position with a grey indicator and no prioritization. Throughout the sample period, the platform underwent interface improvements including deeper integration with the queue and image viewing systems, though the core functionality remained consistent.

\subsection{Data Sources}
\label{sec:data_sources}

We obtained data from three sources: the hospital system's electronic health records, the AI platform vendor, and national provider registries. The electronic health records provide comprehensive clinical data on all CTPA scans, patient characteristics, radiologist identities, and health outcomes. The AI vendor data contain predictions, engagement metrics, and workflow timestamps for all scans processed after system deployment. The national registries provide radiologist demographics information. Together, these sources enable us to link individual patients to their scans, radiologists, AI predictions, and subsequent health outcomes.

\paragraph{Electronic Health Records} First, we extracted CTPA scan records, radiology reports, and patient outcomes from the health system's Observational Medical Outcomes Partnership (OMOP) Common Data Model. OMOP is a standardized healthcare data model that integrates electronic health records across institutions \citep{overhage2012validation}. The OMOP database contains all CTPA scans performed at the health system from May 2012 to August 2024. For each scan, we observe the patient identifier, scan date and time, ordering provider, and care site location. We track the full clinical chain of custody, including preliminary interpreting radiologists (residents or fellows) and signing (attending) radiologists who perform final review and validation. We extracted radiology reports containing radiologist names from OMOP note tables using natural language processing to identify PE diagnoses. 

We extracted patient outcomes from OMOP clinical data, including 16.7 million clinical events spanning 58 outcome types: mortality, emergency visits, hospital readmissions, anticoagulation (blood thinning medication) orders, bleeding events, and cardiovascular complications. For each outcome, we determined the time window from scan to event and constructed binary indicators for outcome occurrence. We also extracted patient control variables to test quasi-random assignment of patients to radiologists. We collected 60 control variables including demographics (age, sex, race, marital status, religion), vital signs (blood pressure, pulse, respiratory rate, temperature, oxygen saturation), clinical presentation (chest pain, shortness of breath, hemoptysis, fever, syncope, tachycardia), laboratory values (D-dimer level, white blood cell count, arterial blood gas analysis), risk factors (current smoking, immobilization, recent surgery, pregnancy, prior PE or DVT, malignancy), healthcare utilization (prior year Emergency Department visits, inpatient admissions, outpatient visits), ordering provider characteristics (total orders placed), and radiological context (number of readers). We extracted 37 comorbidity categories using established clinical classification systems including heart failure, chronic obstructive pulmonary disease, diabetes, hypertension, malignancy, liver disease, renal failure, coagulopathy, obesity, and substance use disorders. The comorbidity mapping uses ICD-10 diagnosis codes to create binary indicators for each condition.

We identify CTPA scans using Epic procedure codes from the hospital's order entry system. Epic is the electronic health record platform used by the hospital system. We include scans ordered with three CTPA codes that indicate suspected PE: IMG1266 (CT CHEST PULMONARY EMBOLISM W IV CONTRAST), IMG3551 (CTA CHEST (PE) ABDOMEN PELVIS W IV CONTRAST), and IMG3904 (CTA CHEST (PE) W IV CONTRAST). These codes identify scans where PE is the primary pathology targeted based on clinical suspicion. We exclude scans ordered for other chest imaging indications where PE detection would be incidental, such as general chest CT, cardiac imaging, or aortic evaluation.

\paragraph{AI Platform Developer} Second, we obtained AI predictions and radiologist engagement metrics from the AI system developer. The vendor provided predictions for all scans processed after system deployment at each site. For each scan, we observe whether or not the AI flagged PE. We track radiologist interaction with the AI system through engagement metrics, including whether the radiologist accessed the AI alert, the type of interaction, engagement duration in minutes, and timestamps of initial engagement. We captured workflow timestamps including when the scan was opened, when the report draft was initiated, and when the final report was signed. The engagement data also contain radiologist names as recorded by the AI system.

\paragraph{National Provider Registries} Third, we obtained radiologist characteristics from national provider databases. We matched radiologist names from the OMOP provider tables to the National Provider Identifier (NPI) registry using name matching via a large language model (LLM). The NPI registry is maintained by the Centers for Medicare \& Medicaid Services and provides unique identifiers for all healthcare providers in the United States. We used the November 2025 NPPES Data Dissemination files \citep{cms2025nppes}. The matching algorithm accounted for name variations including maiden names, nicknames, middle name differences, and common data entry errors. We then linked NPI numbers to CMS physician databases containing medical school graduation year, gender, and practice characteristics. 

\paragraph{Linking Radiologists Across Data Sources} We linked radiologists across the four data sources---OMOP provider records, radiology report signatures, AI engagement data, and NPPES/CMS databases---to create a unified radiologist identification system. We used a multi-stage matching process combining exact matching on provider IDs and NPIs, fuzzy string matching with 85\% similarity thresholds, and AI-assisted disambiguation using an LLM for complex cases. The matching process identified 889 unique name-identifier combinations representing 843 distinct radiologists. We assigned each radiologist a master identifier to handle cases where the same individual appeared under multiple names. The final linked dataset enables tracking individual radiologists across all data sources.

\subsection{Descriptive Statistics}
\label{sec:descriptive_statistics}

\paragraph{Care Sites and Diagnosis Rates}
Table~\ref{tab:care_site_summary} presents summary statistics for the eight care sites in the sample, which are anonymized and sorted by scan volume. The sample includes 117,063 CTPA scans from 84,640 patients. The three largest sites (Sites 1--3) account for 75\% of total scan volume and implemented the AI in August--September 2019. The remaining five sites implemented the system between July 2020 and July 2022. Site volumes range from 200 to 47,175 scans. A total of 388 signing radiologists interpreted scans across all sites that eventually adopted the AI system. The hospital system-wide PE positive diagnosis rate is 10.2\%, with site-level rates ranging from 6.9\% to 11.3\%. Two sites with one scan each were excluded from the analysis, and neither implemented the AI.

\cref{fig:monthly_ctpa_volume,fig:monthly_pe_rate} display monthly scan volume and PE positive diagnosis rates over the full sample period 2012--2024. The orange shaded region indicates the AI rollout period (August 2019--July 2022). Overall, there is a steady increase in the monthly volume over the sample. Monthly scan volume increases from 300 to 900 scans during 2013--2020, with a noticeable decline during early 2020 (COVID-19), then rises to 1,200--1,600 scans during late 2020--2022. PE positivity rates range from 7\% to 14\% with no discernible change before, during, or after the AI rollout period. The stable positivity rate during and after AI deployment suggests the system did not substantially increase detection of new PE cases or change radiologist diagnoses of existing cases, or that these effects offset each other.

\paragraph{Patient Characteristics}
\Cref{tab:patient_characteristics} presents summary statistics for 84,640 patients in the sample who received CTPA scans. The sample is evenly split between the pre-AI period (42,612 patients) and AI period (42,028 patients). The patient population is predominantly female (60\%), White (69\%), and averages 60 years old. Most patients (78\%) receive a single scan. Radiologists diagnose PE in 12\% of patients. Mortality rates are 5\% at 30 days, 7\% at 90 days, and 12\% at one year. Patient demographics remain stable across periods, with differences under 3 percentage points for all groups. The most substantial change is in scan utilization: patients in the AI period are 10 percentage points less likely to receive multiple scans. This reduction suggests that AI may have reduced diagnostic uncertainty, though the overall PE diagnosis rate shows no meaningful change.

\paragraph{Scan Characteristics}
\Cref{tab:scan_characteristics} presents summary statistics for 117,065 CTPA scans across the pre-AI period (54,150 scans) and AI period (62,915 scans). Emergency department scans increase from 56\% to 75\%, an 18 percentage point rise. Repeat imaging within 7 days increases from 1.0\% to 1.5\%, and repeat imaging within 30 days increases from 3.6\% to 5.3\%. Radiologists diagnose PE in 10.11\% of scans in the pre-AI period with virtually no change after the staggered AI rollout (10.25\%, an imprecisely estimated difference). AI flags PE in 10.66\% of scans during the AI period. Radiologists and AI agree in 96.7\% of cases: 8.9\% both positive and 87.8\% both negative. AI detects PE without radiologist confirmation in 1.7\% of cases, while radiologists diagnose PE despite negative AI predictions in 1.6\% of cases. Workflow patterns remain stable, with day shift: 7am-3pm (15--16\%), evening shift: 3pm-11pm (51--52\%), and night shift 11pm-7am (33--34\%) proportions of scans staying essentially unchanged across periods.

\paragraph{Radiologist Characteristics}
\Cref{tab:radiologist_characteristics} presents summary statistics for 389 signing radiologists over the sample period, comparing the pre-AI (309 radiologists) and AI periods (187 radiologists). Panel A shows mean scans per radiologist per month increase from 5.3 to 10.4, while the median increases from 3.2 to 5.4. This near-doubling of monthly reading volume may reflect productivity gains from AI implementation, compositional changes in the radiologist workforce, or secular trends in scan volumes and staffing patterns. Radiologist workload shows large heterogeneity---the 95th percentile radiologist reads 30.3 scans per month compared to 1.0 at the 5th percentile. Panel B shows gender composition remains stable at 32.6\% female in the pre-AI period and 30.4\% in the AI period. Mean years since medical school (i.e., years of experience) increases from 13.3 to 15.8, while mean years active in sample remains similar at 1.7 years in both periods despite the pre-AI period spanning 7.4 years versus 2.2 to 5.2 years for the AI period based on the staggered rollout. Panel C shows the mean PE diagnosis rate at the radiologist level increases from 10.45\% to 12.05\%, with the median increasing from 8.74\% to 10.13\% and the interquartile range narrowing from [0.00\%, 12.24\%] to [7.14\%, 12.97\%].

\paragraph{Scan Efficiency}
As shown in \Cref{tab:scan_efficiency}, diagnostic speed remains stable across periods: mean time from order to diagnosis increases from 3.57 to 3.72 hours (+0.15 hours, se=0.07), even as total scan volume increases 16\% from 54,150 to 62,915 scans (\Cref{tab:care_site_summary}). The stable per-scan reading time alongside higher per-radiologist monthly volumes suggests potential productivity gains operate through optimizing the workflow---such as faster case triage, reduced cognitive load from AI serving as a concurrent second reader, or more efficient handling of negative cases---rather than compressed decision-making time. Reading time for PE-positive cases increases from 3.23 to 3.61 hours (+0.39 hours), consistent with a longer review of AI-flagged findings rather than a rushed diagnosis.

\paragraph{Radiologist-AI (Dis)agreement}
Panel D of \Cref{tab:radiologist_characteristics} reveals distinct differences in radiologist-AI (dis)agreement patterns: radiologists show high consensus when the AI diagnosis is negative ($P(\text{Rad}- | \text{AI}-) = 97.2\%$), while agreement is lower when the AI flags a positive case ($P(\text{Rad}+ | \text{AI}+) = 84.1\%$). This asymmetry suggests radiologists more readily confirm AI-negative predictions than AI-positive predictions, consistent with varying thresholds for accepting AI-flagged findings. The disagreement patterns show that radiologists diagnose PE in 2.8\% of cases where AI predicts no PE, while declining to diagnose PE in 15.9\% of cases where AI predicts PE presence. \Cref{fig:rad_agreement_rates} visualizes this asymmetry in agreement patterns across radiologists, showing a tight, high-density peak near 1.0 for negative AI predictions and a wider, lower-centered distribution for positive AI predictions. This heterogeneity in AI-positive cases reflects varying radiologist thresholds for confirming AI-flagged findings compared to near-universal agreement on scans where AI detects no PE.

\paragraph{Radiologist Engagement with AI}
\Cref{tab:engagement_descriptive} presents radiologist-level engagement patterns with the AI system. The sample includes 148 radiologists who received AI notifications and signed the corresponding reports. Panel A shows radiologists receive a median of 239 AI notifications (IQR: 31 to 946), with mean of 702 (SD = 1,106), reflecting substantial heterogeneity in notification volume. They hover over AI predictions (the key image in the scan that triggered a positive flag) in 32\% of notifications (median), with engagement rates ranging from 21\% to 45\% at the interquartile range. Hover duration is uniformly one minute across the sample. Panel B reveals AI alerts arrive after radiologists have already opened cases in 99.7\% of instances (median), indicating AI functions as a concurrent check rather than advance triage. When AI does alert first, the median triage lead is 8.8 minutes. Radiologists respond to AI alerts within a median of 4.0 minutes. Panel C shows median time from opening to draft is zero minutes, indicating radiologists often begin dictation immediately upon viewing scans. Draft-to-finalization takes a median of 14.7 minutes (IQR: 7.9 to 22.2), accounting for most workflow duration. Total time-to-finalization shows a median of 9.1 minutes but extreme variance (mean = 53.9, SD = 224.2).  Panel D reveals engagement effects on AI-positive cases: radiologists who hover finalize reports in 30.1 minutes on average compared to 52.1 minutes without hover, a mean time savings of 25.7 minutes (SD = 96.4), though the median savings is only 1.4 minutes with wide variation (IQR: -3.1 to 7.9 minutes).

\section{Main Results: Radiologist Diagnoses with AI}
\label{sec:main_results}

This section examines how radiologists collaborate with AI when diagnosing pulmonary embolism. We focus on four key questions. First, how do disagreement patterns between radiologists and AI evolve over time following system deployment? Second, how much heterogeneity exists across radiologists in their willingness to accept or reject AI recommendations? Third, what radiologist characteristics and behaviors associate with different collaboration patterns? Fourth, how do disagreements resolve when patients return for follow-up imaging?

In brief, we document that disagreement rates decline sharply in the first two years after AI deployment, then stabilize. Large heterogeneity persists across radiologists even after five years of use. Female radiologists show higher agreement with AI-detected PEs than male radiologists. Engagement with the AI system exhibits a non-monotonic relationship with agreement: moderate engagement is correlated with the highest agreement with the AI while both low and high engagement show more frequent disagreement. Follow-up imaging reveals asymmetric resolution patterns depending on the direction of initial disagreement.

\paragraph{Disagreement Over Time}
\label{sec:disagreement_time}

Our initial analysis reveals interesting patterns in how disagreement evolves between the AI's predictions and radiologists' diagnoses following the system's implementation. \cref{fig:disagreement_time_all,fig:disagreement_time_ed} show the rate of disagreement over the five years since the system's initial rollout in 2019. The plotted rate of disagreement is conditional on the AI diagnosis, with the left panel showing cases where AI predicts no PE but radiologists diagnose PE, and the right panel showing the reverse---cases where AI detects PE but radiologists do not.

The results reveal two distinct patterns that emerge consistently across both the full sample (\cref{fig:disagreement_time_all}) and the Emergency Department (ED) subset (\cref{fig:disagreement_time_ed}). Scans from the ED are more likely to be acute cases and there may potentially be stronger evidence of quasi-random assignment of patients to reading radiologists, which is why we subset on it. First, the rate of disagremeent when the AI detects PE but radiologists do not (right panel), starts high at nearly 30\% but drops dramatically by approximately 18 percentage points to around 13\% by year two---a 60\% reduction---then roughly stabilizes. Second, when the AI does not detect PE but radiologists do (left panel), disagreement remains consistently low at 2-3\% throughout the period, with minimal variation over time.

The substantial reduction in radiologist disagreement with AI-positive cases could reflect several mechanisms. One possibility is \textit{automation bias}---the tendency for radiologists to gradually over-rely on AI recommendations and reduce their independent critical evaluation over time \citep{parasuraman1997humans}. Alternatively, the convergence could represent beneficial learning, as radiologists observe patient outcomes for AI-positive cases and rationally calibrate their trust in the system's demonstrated accuracy. Institutional incentives that encourage alignment with AI recommendations to reduce missed diagnoses and potential liability may further contribute to this convergence, regardless of whether it represents appropriate trust or excessive reliance.

To understand whether these aggregate patterns mask important heterogeneity, we display in \cref{fig:disagreement_distribution} how disagreement rates vary across different types of radiologists, focusing on radiologist-AI disagreement. The left panel shows disagreement rates when AI predicts PE, while the right panel shows disagreement rates when AI does not predict PE. The plots show how the cross-sectional distribution of disagreement evolves year by year, with radiologists able to move to different parts of the distribution over time. The different lines represent percentiles of the disagreement distribution in each year after rollout over all care sites.

The patterns reveal how the overall distribution of radiologist behavior shifts following AI implementation. In the left panel, the most ``disagreeable'' radiologists in any given year (95th percentile, dotted line) show sharp changes: starting with disagreement rates around 65\% in year zero, dropping to 30\% by year two, then rising back to about 50\% by year five. The 75th and 90th percentiles show similar U-shaped patterns with declines followed by partial rebounds. The median radiologist (50th percentile, dashed line) shows a steady decline from around 20\% to 0\%, while the most agreeable radiologists (25th percentile) maintain stable disagreement rates of 0\% throughout (not displayed).

The right panel shows that even when AI does not predict PE, the distribution of disagreement still evolves over time. The disagreement rates among the radiologists who disagree the least with the AI (25th percentile) remain flat at 0\% (not displayed), while the 50th and 75th percentiles show steady and sharp declines. Only the highest quantiles (90th and 95th percentiles) exhibit U-shaped patterns, with initial declines followed by partial rebounds, perhaps indicating a rebalancing toward greater willingness to disagree and diagnose PE even when AI suggests otherwise.

These distributional changes suggest that the overall population of radiologists experienced systematic behavioral shifts following AI implementation, with the greatest changes occurring among those most inclined to disagree with AI recommendations in any given period. The U-shaped pattern in the upper quantiles may indicate an initial period of excessive deference to AI followed by a shift toward more independent clinical judgment.

\paragraph{Disagreement by Scan Timing}
\label{sec:disagreement_shift}

\Cref{tab:disagreement_shift} presents radiologist-AI agreement patterns stratified by scan timing for 53,880 CTPA scans in the AI period. Panel A compares shifts defined by time of day: Day (7am-3pm), Evening (3pm-11pm), and Night (11pm-7am). Panel B compares weekday versus weekend scans. Statistics are at the scan level.

Conditional disagreement rates show modest variation across shifts. When AI predicts PE, radiologists disagree (diagnose no PE) in 18.5\% of cases during day shifts, 16.3\% during evening shifts, and 15.4\% during night shifts. Day shifts show a 3.1 percentage point higher disagreement rate than night shifts, though a joint F-test does not detect statistically significant differences across shifts ($F=2.10$, $p=0.123$). When AI predicts no PE, radiologists disagree (diagnose PE) in 1.8\% of cases during both day and evening shifts and 1.7\% during night shifts, with no significant variation ($F=0.66$, $p=0.519$). Weekend versus weekday comparisons yield similar results. Radiologists disagree with positive AI predictions 17.0\% of weekends versus 16.1\% of weekdays ($t=0.75$, $p=0.455$), and disagree with negative AI predictions 1.8\% of weekends versus 1.7\% of weekdays ($t=0.74$, $p=0.459$).

Overall disagreement rates combine both types of disagreement weighted by AI prediction prevalence. These rates remain stable at 3.54\% for day shift, 3.39\% for evening shift, and 3.09\% for night shift, with no statistically significant differences ($F=2.28$, $p=0.102$). Weekend scans show a disagreement rate of 3.44\% compared to 3.27\% for weekday scans ($t=0.90$, $p=0.369$). The modest numerical variation in overall disagreement rates reflects both the conditional disagreement probabilities and differences in AI prediction patterns across shifts, neither of which varies significantly.

Night shifts process the highest proportion of Emergency Department scans (83.2\%), followed by weekend scans (78.3\%). Day shifts have the lowest ER proportion (76.5\%). ER and non-ER scans show significant differences in overall agreement patterns: overall disagreement rates are 3.04\% for ER scans versus 4.10\% for non-ER scans ($t=-5.62$, $p<0.001$). However, this difference primarily reflects compositional effects rather than systematic differences in radiologist behavior. When AI predicts PE, radiologists disagree in 15.93\% of ER cases versus 17.18\% of non-ER cases ($t=-1.16$, $p=0.245$). When AI predicts no PE, radiologists agree in 98.39\% of ER cases versus 97.80\% of non-ER cases ($t=3.97$, $p<0.001$). Despite these modest differences in case mix and conditional rates, disagreement patterns remain broadly consistent across timing categories.

\paragraph{Disagreement by Radiologist Demographics}
\label{sec:disagreement_demo}

\Cref{tab:disagreement_demographics} presents radiologist-AI agreement patterns by demographic characteristics for 169 radiologists in the AI period. Statistics are calculated at the radiologist level: each radiologist's conditional agreement rates are computed from their individual scans, then averaged across radiologists within demographic groups, treating each radiologist equally regardless of scan volume. Panel A compares male and female radiologists. Panel B stratifies by experience quartiles measured as years since medical school graduation. Panel C examines scan volume quartiles.

Female radiologists show higher agreement when AI predicts PE: they diagnose PE in 88.4\% of AI-positive cases compared to 82.0\% for male radiologists ($t=-2.11$, $p=0.037$). Put another way, female radiologists are less likely to override positive AI predictions (11.6\% of cases) compared to 18.0\% for male radiologists. When AI predicts no PE, agreement rates are 96.0\% for female radiologists versus 97.8\% for male radiologists ($t=0.81$, $p=0.420$). Hence, the gender difference is concentrated in responses to positive AI flags.

Experience shows no precisely estimated differences across quartiles or in linear trends. When AI predicts PE, agreement rates range from 80.6\% for the most experienced quartile (21+ years) to 87.2\% for the second quartile (6-10 years), with no significant differences across groups ($F=0.79$, $p=0.502$) and no significant linear trend ($\beta=-0.18$ percentage points per year, $p=0.242$). When AI predicts no PE, agreement rates range from 95.6\% for the least experienced quartile to 98.6\% for the third quartile, with no significant differences across groups ($F=0.75$, $p=0.526$) and no significant linear trend ($\beta=0.09$ percentage points per year, $p=0.244$). More experienced radiologists diagnose PE less frequently when AI predicts no PE: rates decline from 4.4\% in the least experienced quartile to 1.4\% in the third quartile, though the linear trend is not significant ($\beta=-0.09$ percentage points per year, $p=0.244$).

Scan volume quartiles also show no precisely estimated differences in conditional agreement rates. When AI predicts PE, agreement ranges from 83.4\% to 85.3\% across volume quartiles with no significant differences ($F=0.09$, $p=0.966$). When AI predicts no PE, agreement ranges from 94.7\% for low-volume radiologists to 98.5\% for the third quartile, with no significant differences across groups ($F=1.59$, $p=0.194$). Low-volume radiologists diagnose PE in 5.3\% of AI-negative cases compared to 1.6\% for high-volume radiologists, but a linear relation is not precisely estimated ($\beta=-0.001$ percentage points per scan, $p=0.177$). Volume quartiles show greater dispersion in scan counts (ranging from median 7 scans in Q1 to median 982 scans in Q4) than experience quartiles, but conditional agreement patterns remain similar across both dimensions.
 
\paragraph{Disagreement by AI System Engagement}
\label{sec:disagreement_engagement}

\Cref{tab:disagreement_engagement} presents radiologist-AI (dis)agreement patterns by quartile of average engagement rate across 153 radiologists in the AI period. Engagement rate is calculated as the proportion of positive PE notifications that the radiologist hovered over before signing the report, averaged across all scans for each radiologist. Statistics are calculated at the radiologist level: each radiologist's conditional agreement rates are computed from their individual scans, then averaged across radiologists within engagement quartiles, treating each radiologist equally irrespective of scan volume.

In Figure \ref{fig:disagreement_engagement}, we plot the disagreement rates across quartiles of engagement for AI positive scans. When the AI predicts PE, agreement rates vary significantly across engagement quartiles ($F=3.10$, $p=0.029$). Quartile 3 radiologists (26.3\% average engagement) show the highest agreement at 90.9\%, while Quartile 1 (8.3\% engagement) and Quartile 4 (44.5\% engagement) both show 80.8--80.9\% agreement. Quartile 2 radiologists (17.8\% engagement) show intermediate agreement at 87.7\%. The non-monotonic pattern yields no linear trend: the regression coefficient is effectively zero ($\beta=-0.03$ percentage points per percentage point increase in engagement, $p=0.840$). Equivalently, disagreement rates when AI predicts PE range from 9.1\% in Quartile 3 to 19.1--19.2\% in Quartiles 1 and 4.

When AI predicts no PE, agreement rates show no differences across engagement quartiles, ranging from 96.0\% to 98.6\% ($F=0.53$, $p=0.661$). No linear trend is detected ($\beta=-0.02$ percentage points per percentage point increase in engagement, $p=0.603$). Radiologists diagnose PE when AI predicts no PE in 1.4\% to 4.0\% of cases across quartiles, with no differences across groups ($F=0.53$, $p=0.661$) and no linear trend ($\beta=0.02$, $p=0.603$).

In sum, the moderate engagement group (Quartile 3) shows higher agreement with positive AI predictions than both low and high engagement groups. Low engagement radiologists (Quartile 1) and high engagement radiologists (Quartile 4) exhibit similar agreement rates when AI predicts PE, despite a fivefold difference in engagement rates (8.3\% versus 44.5\%). Agreement when AI predicts no PE remains high and stable across all engagement levels.

These findings suggest a nuanced relationship between AI engagement and diagnostic alignment between the radiologist and the AI. Two modes of disagreement may exist. Low engagement in the first quartile may represent a ``passive bypass.'' These radiologists likely rely on their own judgment and skip alerts. High disagreement in the fourth quartile suggests an ``active vetting" behavior. By looking at AI evidence more often, these radiologists may find reasons to reject AI errors or artifacts. Radiologists with moderate engagement (Quartile 3) show the highest agreement with the system. This level of interaction may represent a ``sweet spot'' where radiologists use the tool for confirmation without over-scrutinizing every flag. High engagement, by contrast, may lead to a more skeptical view of the system.

\paragraph{Diagnostic Transitions between Sequential Scans}
\label{sec:diagnosis_transition}

When radiologists and AI disagree, what happens next? Do disagreements persist or resolve? When both agree, does subsequent imaging confirm the assessment? 

To study these questions, we examine 35,896 patients who receive an initial CTPA scan with AI assistance. Of these, 34,521 patients (96.2\%) do not receive a second scan within 30 days. The remaining 1,375 patients (3.8\%) receive follow-up imaging. We use a 30-day window to capture short-term diagnostic evolution while excluding routine surveillance scans that typically occur at longer intervals. \Cref{fig:transition_heatmaps} displays transition probabilities from diagnostic states at the first scan (rows) to states at the second scan (columns). Each cell shows the percentage of patients moving between categories. The four diagnostic states are: \textit{Rad Only (+)} where only the radiologist diagnosed PE, \textit{Agree (+)} where both the AI and radiologist detected PE, \textit{AI Only (+)} where only AI flagged PE, and \textit{Agree (-)} where both found no PE.

We test whether transition patterns differ from chance. Let $O_{ij}$ denote observed transitions from state $i$ to state $j$. Let $E_{ij}$ denote expected transitions under independence. The standardized residual is:
\begin{equation}
t_{ij} = \frac{O_{ij} - E_{ij}}{\sqrt{E_{ij}(1-r_i)(1-c_j)}}
\end{equation}
where $r_i$ is the proportion in starting state $i$ and $c_j$ is the proportion in ending state $j$. Values exceeding $|t_{ij}| > 1.96$ indicate transitions that occur more or less often than expected. For comparing two proportions $p_1$ and $p_2$, the test statistic is:
\begin{equation}
t = \frac{p_1 - p_2}{\sqrt{\hat{p}(1-\hat{p})(1/N_1 + 1/N_2)}}
\end{equation}
where $\hat{p}$ is the pooled proportion.

\Cref{fig:transition_heatmap_unconditional} shows unconditional probabilities for all 35,896 patients with a first scan, including those who exit the sample. Exit rates range from 90.6\% for \textit{Rad Only (+)} to 96.5\% for \textit{Agree (-)}. The highest exit rate occurs when both AI and radiologist agree on no PE, likely reflecting clinical confidence that no further imaging is needed. The lowest exit rate occurs when only the radiologist diagnoses PE. This 5.9 percentage point difference suggests that when radiologists override AI to diagnose PE, more follow-up scans are ordered ($t=2.78$).

Among the small fraction with follow-up scans, non-Exit transitions appear modest: 0.1\% to 5.1\%. Some states show strong persistence. \textit{Rad Only (+)} to \textit{Rad Only (+)} has $t=16.76$. \textit{Agree (+)} to \textit{Agree (+)} has $t=14.34$. Among all non-exit transitions, 75.3\% maintain the same state versus 24.7\% that change ($t=26.51$). Concordant diagnoses exit 96.3\% of the time. Discordant cases exit 92.8\% ($t=5.69$). Agreement appears to reduce follow-up imaging while disagreement introduces uncertainty or signals more complex cases that require follow-up imaging. Initial disagreement appears to also create friction: discordant cases remain discordant 20.3\% of the time versus 6.6\% for concordant cases becoming discordant ($t=4.38$).

\Cref{fig:transition_heatmap_conditional} shows conditional probabilities among the 1,375 patients who return for follow-up imaging, excluding the Exit state. This reveals diagnostic transitions masked by high exit rates. In the unconditional view, non-Exit transitions appear small (0.1\% to 5.1\%). In the conditional view, these same transitions represent 1.7\% to 87.1\% of follow-up cases.

Three patterns emerge. First, persistence varies by diagnostic state. Among patients with follow-up scans, 75.4\% remain in the same state ($t=26.71$). \textit{Agree (-)} shows the highest persistence at 87.1\%. \textit{Rad Only (+)} persists at 31.2\%. \textit{Agree (+)} persists at 36.1\%. \textit{AI Only (+)} shows zero persistence. Interestingly, when radiologists initially reject AI flags, no follow-up scan shows the radiologist changing to agree with the AI that maintains the positive diagnosis. Concordant negatives are more stable than concordant positives: 87.1\% of \textit{Agree (-)} cases persist versus 36.1\% of \textit{Agree (+)} cases ($t=-16.76$).

Second, when radiologists override AI to diagnose PE, most follow-up scans show resolution. Among \textit{Rad Only (+)} cases, 68.8\% transition to other states. The largest transition is to \textit{Agree (-)} at 54.2\%. Here, the radiologist initially diagnosed PE despite negative AI, but follow-up imaging shows both agreeing on no PE. In contrast, when radiologists initially reject AI flags (\textit{AI Only (+)}), 100\% of follow-up cases transition to other states: 65.4\% to \textit{Agree (-)} and 30.8\% to \textit{Agree (+)}. When disagreements at scan 1 resolve to positive diagnoses at scan 2, the rates are similar: 30.8\% of \textit{AI Only (+)} cases transition to \textit{Agree (+)} versus 14.6\% of \textit{Rad Only (+)} cases ($t=1.65$).

Third, radiologist overrides show different persistence patterns. Among \textit{Rad Only (+)} cases with follow-up, 31.2\% remain \textit{Rad Only (+)}---the radiologist maintains the PE diagnosis across both scans despite AI disagreement. Zero \textit{AI Only (+)} cases remain \textit{AI Only (+)}---when radiologists initially reject AI flags, no follow-up scan shows the pattern persisting ($t=-3.19$). When measuring whether each ``system'' changes its own diagnosis from positive to negative or vice versa between scans, the rates are similar: radiologists change 20.2\% of the time, AI changes 18.6\% ($t=1.06$). The difference lies in which disagreements persist, not overall diagnostic instability.

These patterns reveal how radiologists incorporate AI signals. When radiologists initially reject AI flags, all follow-up scans show state changes---predominantly to \textit{Agree (-)} (65.4\%) rather than \textit{Agree (+)} (30.8\%). When radiologists override negative AI detections to diagnose PE positive, 68.8\% of follow-up scans show state changes, with 54.2\% transitioning to \textit{Agree (-)}. Only 30.8\% of initially rejected AI flags later show concordant positive diagnoses. But 31.2\% of radiologist overrides persist across scans. The 54.2\% transition from \textit{Rad Only (+)} to \textit{Agree (-)} suggests diagnostic uncertainty in some overrides, yet 31.2\% persist. This indicates radiologists incorporate AI signals when revising diagnoses but maintain independent judgment in sustained overrides.

\section{Conclusion}
\label{sec:conclusion}

This paper examines how radiologists collaborate with AI when diagnosing pulmonary embolism. We track 117,063 CTPA scans interpreted by 389 radiologists over twelve years at a large hospital system. The staggered AI rollout between 2019 and 2022 provides variation in exposure timing. We link scans to AI predictions, radiologist diagnoses, engagement metrics, and follow-up imaging.

Five findings emerge. First, radiologists show asymmetric agreement with AI. They agree in 97\% of AI-negative cases but only 84\% of AI-positive cases. This 13 percentage point gap reveals that radiologists more readily accept AI-negative assessments than AI-positive findings. Second, disagreement patterns evolve substantially after deployment. Radiologists reject AI-detected PEs in 30\% of cases in the first year. This drops to 12\% by year two, then stabilizes. The reduction could reflect learning or automation bias. Third, diagnostic efficiency remains stable despite increased workload. Mean time per scan holds at 3.6 hours even as scan volume increases 16\% and per-radiologist monthly volumes double from 5.3 to 10.4 scans. Meanwhile, patient mortality remains roughly the same. The stable per-scan time alongside higher throughput suggests the AI system improves the radiological workflow rather than compromising decision-making. Fourth, large heterogeneity persists across radiologists. Female radiologists are 6 percentage points less likely to override AI-detected PEs than male radiologists. Engagement shows a non-monotonic relationship with agreement: moderate engagement correlates with highest agreement with the AI while both low and high engagement show more disagreement. Fifth, follow-up imaging reveals asymmetric resolution. When radiologists override AI to diagnose PE, 54\% of subsequent scans show both agreeing on no PE. When radiologists reject AI flags, none maintain that pattern.

Overall, radiologists and AI collaborate in predictable but heterogeneous ways. Disagreement declines sharply after deployment but substantial variation persists. Diagnostic speed remains stable despite doubled workload per radiologist. These patterns reveal how professionals incorporate algorithmic recommendations in high-stakes decisions. The asymmetric agreement, time evolution, and persistent heterogeneity suggest that collaboration involves active judgment rather than passive acceptance or rejection.

Our ongoing work exploits the staggered rollout and quasi-random assignment of patients to radiologists to estimate the causal effects of the AI assistance. We will examine the effects of AI on radiologist skill at diagnosing suspected pulmonary embolisms, radiologist preferences for balancing true positive versus false positive rates, and patient health outcomes including mortality, complications, PE treatment, and healthcare utilization like length of hospital stay. Understanding these relationships will inform optimal human-AI collaboration in high-stakes medical decisions.

\pagebreak

\end{spacing}

\singlespacing

\setlength{\bibsep}{0pt plus 0.3ex} 
\fontsize{11}{12}\selectfont 

\bibliographystyle{aer}
\addcontentsline{toc}{section}{\refname}
\bibliography{ref}

@article{duffett2020pulmonary,
  title={Pulmonary embolism: Update on management and controversies},
  author={Duffett, Lisa and Castellucci, Lana A and Forgie, Melissa A},
  journal={BMJ},
  volume={370},
  year={2020},
  publisher={British Medical Journal Publishing Group}
}

@article{freund2022acute,
  title={Acute pulmonary embolism: A review},
  author={Freund, Yonathan and Cohen-Aubart, Fleur and Bloom, Ben},
  journal={JAMA},
  volume={328},
  number={13},
  pages={1336--1345},
  year={2022},
  publisher={American Medical Association}
}

@article{tapson2005acute,
  title={Acute pulmonary embolism},
  author={Tapson, Victor F},
  journal={Management of Acute Decompensated Heart Failure},
  pages={285--300},
  year={2005},
  publisher={CRC Press}
}

@article{grenier2023deep,
  title={Deep learning-based algorithm for automatic detection of pulmonary embolism in chest CT angiograms},
  author={Grenier, Philippe A and Ayobi, Ali and Quenet, Simon and Tassy, Marie and Marx, Maxime and Chow, Daniel S and Weinberg, Bennett D and Chang, Peter D and Chaibi, Yasmina},
  journal={Diagnostics},
  volume={13},
  number={7},
  pages={1324},
  year={2023},
  publisher={MDPI}
}

@article{schoepf2004spiral,
  title={Spiral computed tomography for acute pulmonary embolism},
  author={Schoepf, U Joseph and Goldhaber, Samuel Z and Costello, Philip},
  journal={Circulation},
  volume={109},
  number={18},
  pages={2160--2167},
  year={2004},
  publisher={Am Heart Assoc}
}

@article{weiss2006ct,
  title={CT pulmonary angiography is the first-line imaging test for acute pulmonary embolism: A survey of US clinicians},
  author={Weiss, Clifford R and Scatarige, John C and Diette, Gregory B and Haponik, Edward F and Merriman, Barry and Fishman, Elliot K},
  journal={Academic Radiology},
  volume={13},
  number={4},
  pages={434--446},
  year={2006},
  publisher={Elsevier}
}

@article{anderson2007computed,
  title={Computed tomographic pulmonary angiography vs ventilation-perfusion lung scanning in patients with suspected pulmonary embolism: a randomized controlled trial},
  author={Anderson, David R and Kahn, Susan R and Rodger, Marc A and Kovacs, Michael J and Morris, Tim and Hirsch, Andrew and Lang, Eddy and Stiell, Ian and Kovacs, George and Dreyer, Jon and others},
  journal={JAMA},
  volume={298},
  number={23},
  pages={2743--2753},
  year={2007},
  publisher={American Medical Association}
}

@article{soffer2021deep,
  title={Deep learning for pulmonary embolism detection on computed tomography pulmonary angiogram: A systematic review and meta-analysis},
  author={Soffer, Shelly and Klang, Eyal and Shimon, Orit and Barash, Yiftach and Cahan, Noa and Greenspana, Hayit and Konen, Eli},
  journal={Scientific Reports},
  volume={11},
  number={1},
  pages={15814},
  year={2021},
  publisher={Nature Publishing Group UK London}
}

@article{di2016deep,
  title={Deep vein thrombosis and pulmonary embolism},
  author={Di Nisio, Marcello and van Es, Nick and B{\"u}ller, Harry R},
  journal={The Lancet},
  volume={388},
  number={10063},
  pages={3060--3073},
  year={2016},
  publisher={Elsevier}
}

@article{huhtanen2022automated,
  title={Automated detection of pulmonary embolism from CT-angiograms using deep learning},
  author={Huhtanen, Heidi and Nyman, Mikko and Mohsen, Tarek and Virkki, Arho and Karlsson, Antti and Hirvonen, Jussi},
  journal={BMC Medical Imaging},
  volume={22},
  number={1},
  pages={43},
  year={2022},
  publisher={Springer}
}

@article{weikert2020automated,
  title={Automated detection of pulmonary embolism in CT pulmonary angiograms using an AI-powered algorithm},
  author={Weikert, Thomas and Winkel, David J and Bremerich, Jens and Stieltjes, Bram and Parmar, Victor and Sauter, Alexander W and Sommer, Gregor},
  journal={European Radiology},
  volume={30},
  pages={6545--6553},
  year={2020},
  publisher={Springer}
}

@article{parasuraman1997humans,
  title={Humans and automation: Use, misuse, disuse, abuse},
  author={Parasuraman, Raja and Riley, Victor},
  journal={Human Factors},
  volume={39},
  number={2},
  pages={230--253},
  year={1997},
  doi={10.1518/001872097778543886}
}

@article{cheikh2022artificial,
  title={How artificial intelligence improves radiological interpretation in suspected pulmonary embolism},
  author={Cheikh, Alexandre Ben and Gorincour, Guillaume and Nivet, Hubert and May, Julien and Seux, Mylene and Calame, Paul and Thomson, Vivien and Delabrousse, Eric and Cromb{\'e}, Amandine},
  journal={European Radiology},
  volume={32},
  number={9},
  pages={5831--5842},
  year={2022},
  publisher={Springer}
}

@article{rothenberg2023prospective,
  title={Prospective evaluation of AI triage of pulmonary emboli on CT pulmonary angiograms},
  author={Rothenberg, Steven A and Savage, Cody H and Abou Elkassem, Asser and Singh, Satinder and Abozeed, Mostafa and Hamki, Omar and Junck, Kevin and Tridandapani, Srini and Li, Mei and Li, Yufeng and others},
  journal={Radiology},
  volume={309},
  number={1},
  pages={e230702},
  year={2023},
  publisher={Radiological Society of North America}
}

@article{ebrahimian2022predictive,
  title={Predictive values of AI-based triage model in suboptimal CT pulmonary angiography},
  author={Ebrahimian, Shadi and Digumarthy, Subba R and Homayounieh, Fatemeh and Bizzo, Bernardo C and Dreyer, Keith J and Kalra, Mannudeep K},
  journal={Clinical Imaging},
  volume={86},
  pages={25--30},
  year={2022},
  publisher={Elsevier}
}

@article{liu2019comparison,
  title={A comparison of deep learning performance against health-care professionals in detecting diseases from medical imaging: a systematic review and meta-analysis},
  author={Liu, Xiaoxuan and Faes, Livia and Kale, Aditya U and Wagner, Siegfried K and Fu, Dun Jack and Bruynseels, Alice and Mahendiran, Thushara and Moraes, Gabriella and Shamdas, Mohith and Kern, Christoph and others},
  journal={The Lancet Digital Health},
  volume={1},
  number={6},
  pages={e271--e297},
  year={2019},
  publisher={Elsevier}
}

@article{rajpurkar2023current,
  title={The current and future state of AI interpretation of medical images},
  author={Rajpurkar, Pranav and Lungren, Matthew P},
  journal={New England Journal of Medicine},
  volume={388},
  number={21},
  pages={1981--1990},
  year={2023},
  publisher={Mass Medical Soc}
}

@article{schmuelling2021deep,
  title={Deep learning-based automated detection of pulmonary embolism on CT pulmonary angiograms: No significant effects on report communication times and patient turnaround in the emergency department nine months after technical implementation},
  author={Schmuelling, Lena and Franzeck, Fabian C and Nickel, Christian H and Mansella, Gregory and Bingisser, Roland and Schmidt, Noemi and Stieltjes, Bram and Bremerich, Jens and Sauter, Alexander W and Weikert, Thomas and Gregor Sommer},
  journal={European Journal of Radiology},
  volume={141},
  pages={109816},
  year={2021},
  publisher={Elsevier}
}

@article{angelova2025algorithmic,
  title={Algorithmic recommendations and human discretion},
  author={Angelova, Victoria and Dobbie, Will and Yang, Crystal S},
  journal={Review of Economic Studies},
  year={2025},
  publisher={Oxford University Press UK}
}

@article{dattner2019legal,
  title={The legal and ethical implications of using AI in hiring},
  author={Dattner, Ben and Chamorro-Premuzic, Tomas and Buchband, Richard and Schettler, Lucinda},
  journal={Harvard Business Review},
  volume={25},
  pages={1--7},
  year={2019}
}

@article{Roy2025,
  author  = {Roy, Prasenjit and Ghose, Biswajit and Singh, Premendra Kumar and others},
  title   = {Artificial Intelligence and Finance: A bibliometric review on the Trends, Influences, and Research Directions},
  journal = {F1000Research},
  year    = {2025},
  volume  = {14},
  pages   = {122},
  doi     = {10.12688/f1000research.160959.1},
  url     = {https://f1000research.com/articles/14-122}
}

@unpublished{agarwal2023combining,
  title={Combining human expertise with artificial intelligence: Experimental evidence from radiology},
  author={Agarwal, Nikhil and Moehring, Alex and Rajpurkar, Pranav and Salz, Tobias},
  note={Working Paper},
  year={2025},
  month={November}
}

@Unpublished{harris2024decision,
  author  = {Harris, Adam and Yellen, Maggie},
  note    = {Working Paper},
  title   = {Decision-making with machine prediction: Evidence from predictive maintenance in trucking},
  year    = {2024},
  month   = {January}
}

@article{patel2019human,
  title={Human--machine partnership with artificial intelligence for chest radiograph diagnosis},
  author={Patel, Bhavik N and Rosenberg, Louis and Willcox, Glen and Baltaxe, David and Lyons, Mimi and Irvin, Jeremy and Rajpurkar, Pranav and Amrhein, Timothy and Gupta, Ritu and Halabi, Safwan and Curtis Langlotz and Edward Lo and Joseph Mammarappallil and A. J. Mariano and Geoffrey Riley and Jayne Seekins and Luyao Shen and Evan Zucker and Matthew P. Lungren},
  journal={NPJ Digital Medicine},
  volume={2},
  number={1},
  pages={111},
  year={2019},
  publisher={Nature Publishing Group}
}

@article{chen2024impact,
  title={Impact of human and artificial intelligence collaboration on workload reduction in medical image interpretation},
  author={Chen, Mingyang and Wang, Yuting and Wang, Qiankun and Shi, Jingyi and Wang, Huike and Ye, Zichen and Xue, Peng and Qiao, Youlin},
  journal={NPJ Digital Medicine},
  volume={7},
  number={1},
  pages={349},
  year={2024},
  publisher={Nature Publishing Group UK London}
}

@article{shin2023impact,
  title={The impact of artificial intelligence on the reading times of radiologists for chest radiographs},
  author={Shin, Hyun Joo and Han, Kyunghwa and Ryu, Luke and Kim, Eun-Kyung},
  journal={NPJ Digital Medicine},
  volume={6},
  number={1},
  pages={82},
  year={2023},
  publisher={Nature Publishing Group}
}

@article{leibig2022combining,
  title={Combining the strengths of radiologists and AI for breast cancer screening: a retrospective analysis},
  author={Leibig, Christian and Brehmer, Moritz and Bunk, Stefan and Byng, Daniel and Pinker, Katja and Umutlu, Lale},
  journal={The Lancet Digital Health},
  volume={4},
  number={7},
  pages={e507--e519},
  year={2022},
  publisher={Elsevier}
}

@article{brynjolfsson2017machine,
  title={What can machine learning do? Workforce implications},
  author={Brynjolfsson, Erik and Mitchell, Tom},
  journal={Science},
  volume={358},
  number={6370},
  pages={1530--1534},
  year={2017},
  publisher={American Association for the Advancement of Science}
}

@article{acemoglu2018automation,
  title={Automation and new tasks: How technology displaces and reinstates labor},
  author={Acemoglu, Daron and Restrepo, Pascual},
  journal={Journal of Economic Perspectives},
  volume={33},
  number={2},
  pages={3--30},
  year={2019},
  publisher={American Economic Association}
}

@book{daugherty2018human,
  title={Human + Machine: Reimagining Work in the Age of AI},
  author={Daugherty, Paul R and Wilson, H James},
  year={2024},
  publisher={Harvard Business Review Press}
}

@misc{who2021ethics,
  title={Ethics and governance of artificial intelligence for health: WHO guidance},
  author={{World Health Organization}},
  year={2021},
  publisher={World Health Organization},
  url={https://www.who.int/publications/i/item/9789240037403}
}

@book{shortliffe2015clinical,
  title={Clinical Decision Support: The Road to Broad Adoption},
  author={Shortliffe, Edward H and Cimino, James J},
  year={2014},
  publisher={Elsevier},
  edition={2nd}
}

@article{kelly2019key,
  title={Key challenges for delivering clinical impact with artificial intelligence},
  author={Kelly, Christopher J and Karthikesalingam, Alan and Suleyman, Mustafa and Corrado, Greg and King, Dominic},
  journal={BMC Medicine},
  volume={17},
  number={1},
  pages={1--9},
  year={2019},
  publisher={BioMed Central}
}

@article{geis2019ethics,
  title={Ethics of artificial intelligence in radiology: Summary of the joint European and North American multisociety statement},
  author={Geis, J Raymond and Brady, Adrian P and Wu, Carol C and Spencer, Judy and Ranschaert, Erik and Jaremko, Jacob L and Langer, Steve G and Kitts, Adam Boyce and Birch, Judy and Shields, William F and others},
  journal={Radiology},
  volume={293},
  number={2},
  pages={436--440},
  year={2019},
  publisher={Radiological Society of North America}
}

@article{overhage2012validation,
  title={Validation of a common data model for active safety surveillance research},
  author={Overhage, J Marc and Ryan, Patrick B and Reich, Christian G and Hartzema, Abraham G and Stang, Paul E},
  journal={Journal of the American Medical Informatics Association},
  volume={19},
  number={1},
  pages={54--60},
  year={2012},
  publisher={Oxford University Press}
}

@article{heit2015epidemiology,
  title={Epidemiology of venous thromboembolism},
  author={Heit, John A},
  journal={Nature Reviews Cardiology},
  volume={12},
  number={8},
  pages={464--474},
  year={2015},
  publisher={Nature Publishing Group}
}

@article{torbicki2008guidelines,
  title={Guidelines on the diagnosis and management of acute pulmonary embolism},
  author={Torbicki, Adam and Perrier, Arnaud and Konstantinides, Stavros and Giancarlo Agnelli and Nazzareno Galie` and Piotr Pruszczyk and Frank Bengel and Adrian J.B. Brady and Daniel Ferreira and Uwe Janssen and Walter Klepetko and Eckhard Mayer and Martine Remy-Jardin and Jean-Pierre Bassand},
  journal={European Heart Journal},
  volume={29},
  number={18},
  pages={2276--2315},
  year={2008},
  publisher={Oxford University Press}
}

@article{anderson2003risk,
  title={Risk factors for venous thromboembolism},
  author={Anderson, Frederick A and Spencer, Frederick A},
  journal={Circulation},
  volume={107},
  number={23},
  pages={I--9},
  year={2003},
  publisher={Am Heart Assoc}
}

@article{dentali2010prevalence,
  title={Prevalence and clinical history of incidental, asymptomatic pulmonary embolism: a meta-analysis},
  author={Dentali, Francesco and Ageno, Walter and Becattini, Cecilia and Galli, Luca and Gianni, Marco and Riva, Nicola and Imberti, Davide and Squizzato, Alessandro and Venco, Achille and Agnelli, Giancarlo},
  journal={Thrombosis Research},
  volume={125},
  number={6},
  pages={518--522},
  year={2010},
  publisher={Elsevier}
}

@article{kearon2003natural,
  title={Natural history of venous thromboembolism},
  author={Kearon, Clive},
  journal={Circulation},
  volume={107},
  number={23\_suppl\_1},
  pages={I--22},
  year={2003},
  publisher={Lippincott Williams \& Wilkins}
}

@article{gladish2006incidental,
  title={Incidental pulmonary emboli in oncology patients: prevalence, CT evaluation, and natural history},
  author={Gladish, Gregory W and Choe, Du Hwan and Marom, Edith M and Sabloff, Bradley S and Broemeling, Lyle D and Munden, Reginald F},
  journal={Radiology},
  volume={240},
  number={1},
  pages={246--255},
  year={2006},
  publisher={Radiological Society of North America}
}

@article{o2015treat,
  title={How I treat incidental pulmonary embolism},
  author={O’Connell, Casey},
  journal={Blood, The Journal of the American Society of Hematology},
  volume={125},
  number={12},
  pages={1877--1882},
  year={2015},
  publisher={American Society of Hematology Washington, DC}
}

@misc{cms2025nppes,
  author = {{Centers for Medicare \& Medicaid Services}},
  title = {{NPPES Data Dissemination}},
  year = {2025},
  month = {November},
  note = {Accessed January 2026},
  howpublished = {\href{https://download.cms.gov/nppes/NPI_Files.html}{https://download.cms.gov/nppes/NPI\_Files.html}}
}

\clearpage


\renewcommand{\thefigure}{\arabic{figure}}
\renewcommand{\thesubfigure}{\Alph{subfigure}}
\renewcommand{\thetable}{\arabic{table}}

\captionsetup[figure]{textfont={sc,md},labelformat=simple, labelsep=newline}
\captionsetup[subfigure]{labelfont=bf,textfont=bf,labelformat=parens, labelsep=quad}

\captionsetup[table]{textfont={sc,md},labelformat=simple, labelsep=newline}



\clearpage
\begin{figure}[!t]	
    \centering
    \includegraphics[width=.7\textwidth]{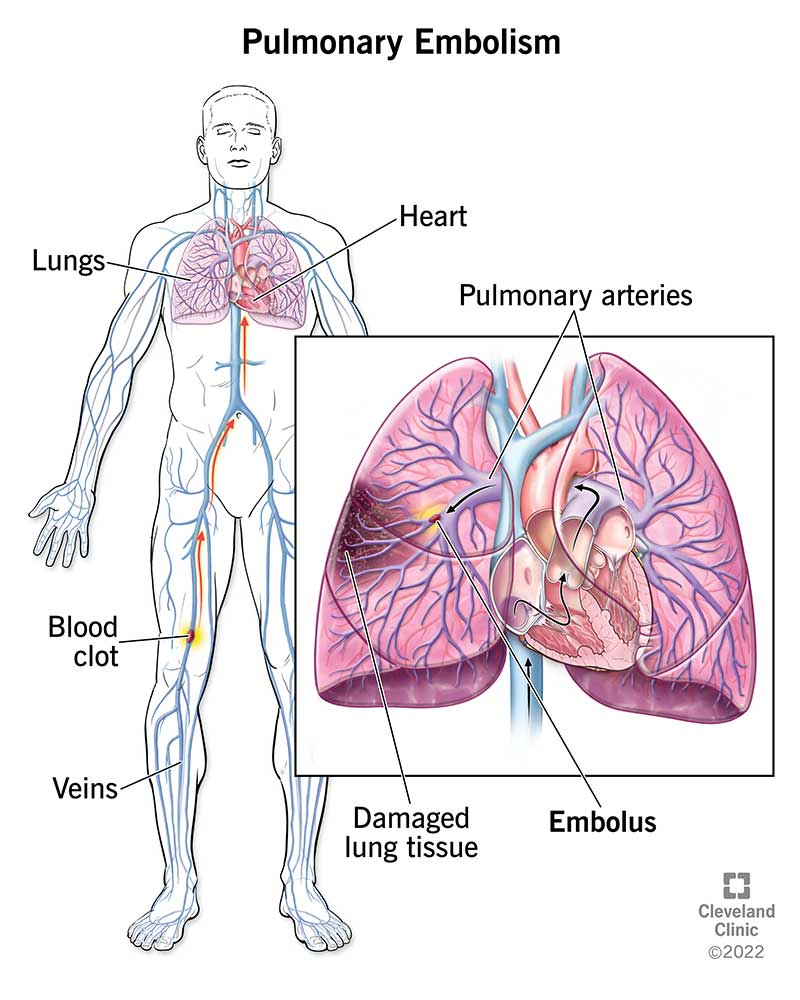}
    \caption{Pulmonary Embolism Anatomy}
    \label{fig:pe_anatomy}
    \medskip
    \begin{minipage}{\textwidth}
    \setlength{\parindent}{0pt}
    The figure illustrates pulmonary embolism pathophysiology. Oftentimes, blood clots form in leg or pelvic veins, travel through the circulatory system, and lodge in pulmonary arteries. The inset shows the embolus blocking blood flow in the lung, causing tissue damage. Image courtesy of \href{https://my.clevelandclinic.org/health/diseases/17400-pulmonary-embolism}{Cleveland Clinic}.
    \end{minipage}
\end{figure}

\clearpage
\begin{figure}[!t]
    \centering
    
    \begin{subfigure}[b]{0.48\textwidth}
        \centering
        \includegraphics[width=\textwidth]{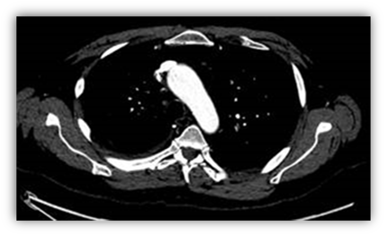}
        \caption{Direct Sign: Pulmonary Artery Filling Defect}
        \label{fig:ctpa_clot}
    \end{subfigure}
    \hfill
    \begin{subfigure}[b]{0.48\textwidth}
        \centering
        \includegraphics[width=\textwidth]{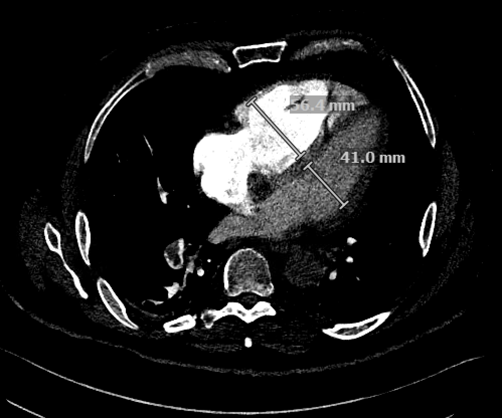}
        \caption{Indirect Sign: Right Heart Strain}
        \label{fig:ctpa_strain}
    \end{subfigure}
    
    \caption{Computed Tomographic Pulmonary Angiography (CTPA) of Pulmonary Embolism}
    \label{fig:ctpa_examples}
    
    \vspace{0.2cm}
    
    \begin{minipage}{\textwidth}
    \setlength{\parindent}{0pt}\footnotesize
    The figure shows two CTPA images from patients with PE. Each image shows a horizontal slice through the chest, as if looking upward from the fee. Panel A shows the upper chest where the main pulmonary artery splits into left and right branches. A grey area appears inside the right pulmonary artery (left side of image) where the vessel should be uniformly bright white from contrast dye. This grey area is the blood clot blocking blood flow. Panel B shows a lower slice through the heart chambers. The right ventricle measures 56.4 mm compared to 41.0 mm for the left ventricle, producing an RV/LV ratio of 1.37. Normally the left ventricle is larger (RV/LV < 1). The enlarged right ventricle indicates the heart is straining to pump blood through the blocked lung vessels. Together, the images illustrate both the direct evidence of PE (the clot itself in Panel A) and its cardiac consequence (right heart strain in Panel B). De-identified CT images courtesy of the hospital system studied.
    \end{minipage}
\end{figure}

\clearpage
\begin{figure}[!t]	
    \centering
    \includegraphics[width=.9\textwidth]{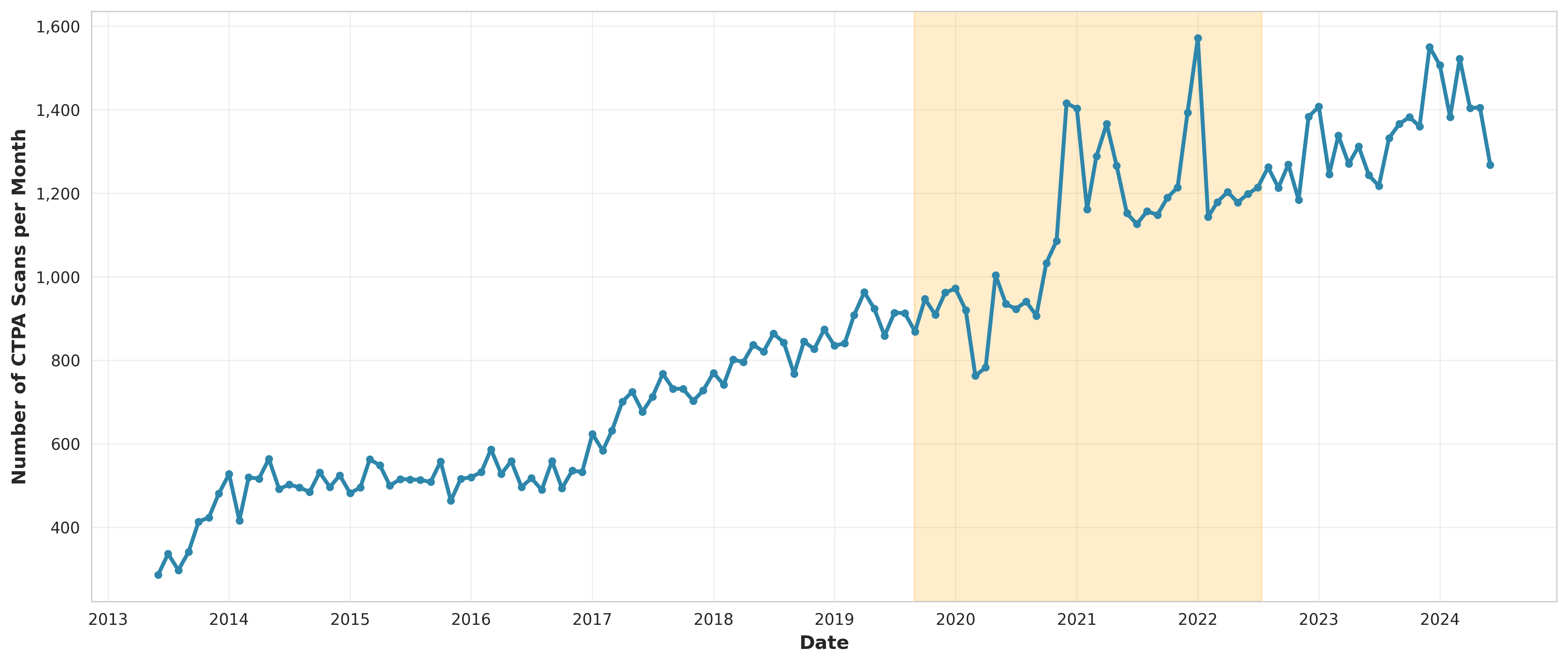}
    \caption{Monthly CTPA Scan Volume}
    \label{fig:monthly_ctpa_volume}
    \medskip
    \begin{minipage}{\textwidth}
    \setlength{\parindent}{0pt}
    The figure shows the monthly volume of CTPA scans across all eight care sites that received the AI system. The blue line represents the number of scans performed each month. The orange shaded region indicates the AI rollout period from 08/29/2019 to 07/12/2022, during which the AI system was progressively deployed across sites. 
    \end{minipage}
\end{figure}

\clearpage
\begin{figure}[!t]	
    \centering
    \includegraphics[width=.9\textwidth]{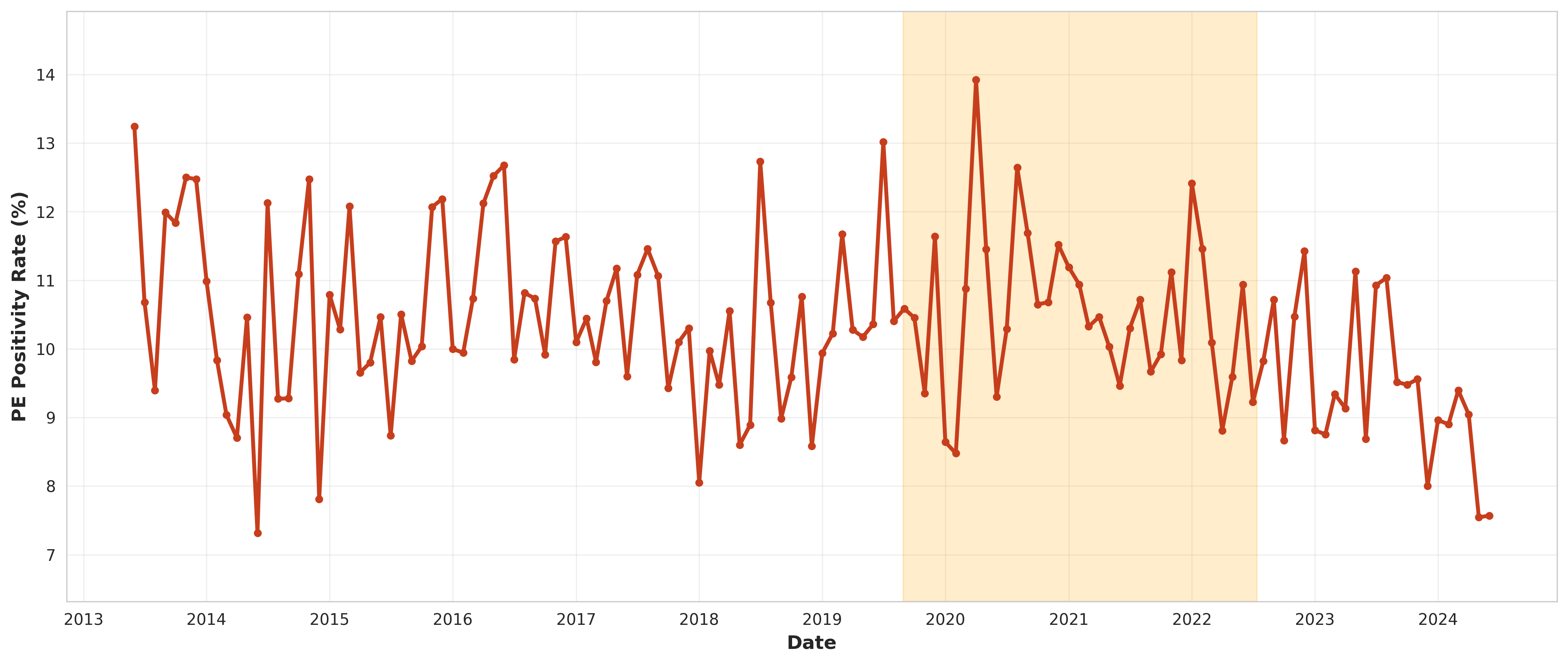}
    \caption{Monthly PE Positivity Rate}
    \label{fig:monthly_pe_rate}
    \medskip
    \begin{minipage}{\textwidth}
    \setlength{\parindent}{0pt}
    The figure shows the monthly PE positive diagnosis rate across all eight care sites that received the AI system. The red line represents the percentage of CTPA scans with a positive PE diagnosis by the signing radiologist each month. The orange shaded region indicates the AI rollout period from 08/29/2019 to 07/12/2022, during which the AI system was progressively deployed across sites.
    \end{minipage}
\end{figure}

\clearpage
\begin{figure}[!t]	
    \centering
    \includegraphics[width=.9\textwidth]{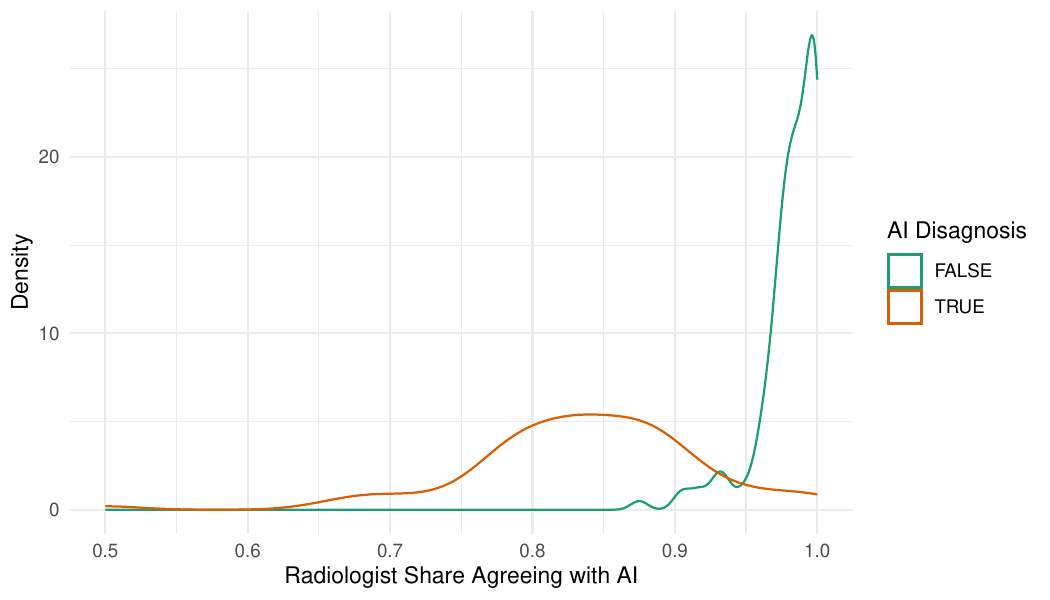}
    \caption{Radiologist Agreement Rates with AI by AI Diagnosis}
    \label{fig:rad_agreement_rates}
    \medskip
    \begin{minipage}{\textwidth}
    \setlength{\parindent}{0pt}
    The figure displays the density distribution of the share of radiologists agreeing with the AI prediction, stratified by the AI's diagnosis. The green line (FALSE) represents cases where the AI did not detect PE. The orange line (TRUE) represents cases where the AI flagged a positive PE.
    \end{minipage}
\end{figure}

\clearpage
\begin{figure}[!t]	
    \centering
    \includegraphics[width=.9\textwidth]{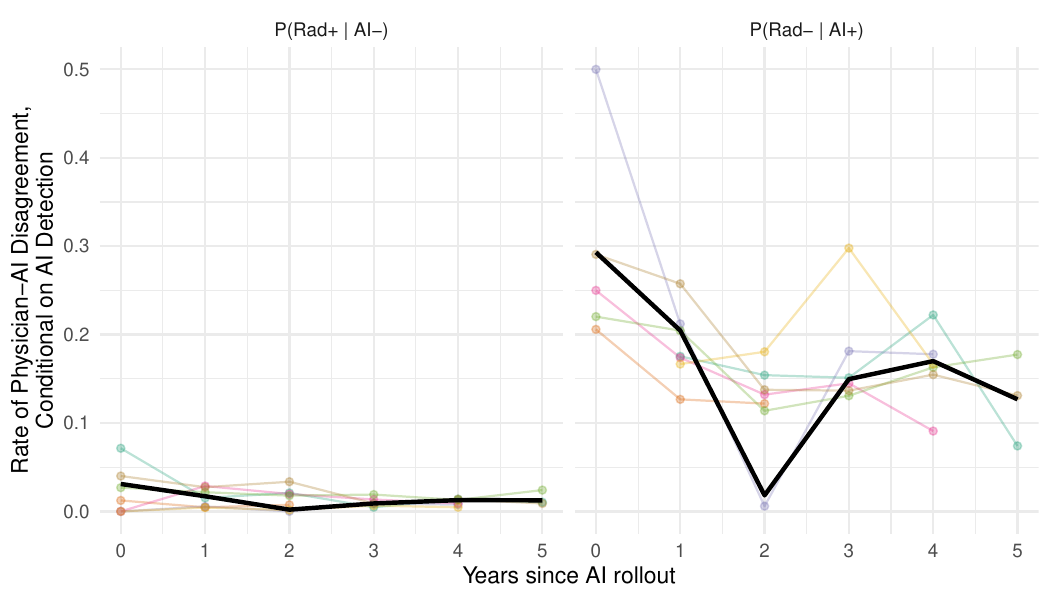}
    \caption{Disagreement Rates Between AI and Radiologists Over Time}
    \label{fig:disagreement_time_all}
    \medskip
    \begin{minipage}{\textwidth}
    \setlength{\parindent}{0pt}
    This figure shows the rate of disagreement between AI predictions and radiologist diagnoses based on all CTPA scans in the sample. The left panel shows disagreement when the AI detects no PE but radiologists observe one. The right panel shows disagreement when the AI detects PE and the radiologist does not. The colored lines are disagreement patterns by care site per the staggered rollout of the AI system. The solid black lines are simple average per period. Time is measured in years since the initial AI rollout in 2019 per \cref{tab:care_site_summary}. Care site 8 is excluded due to its limited scan volume.
    \end{minipage}
\end{figure}

\clearpage
\begin{figure}[!t]	
    \centering
    \includegraphics[width=.9\textwidth]{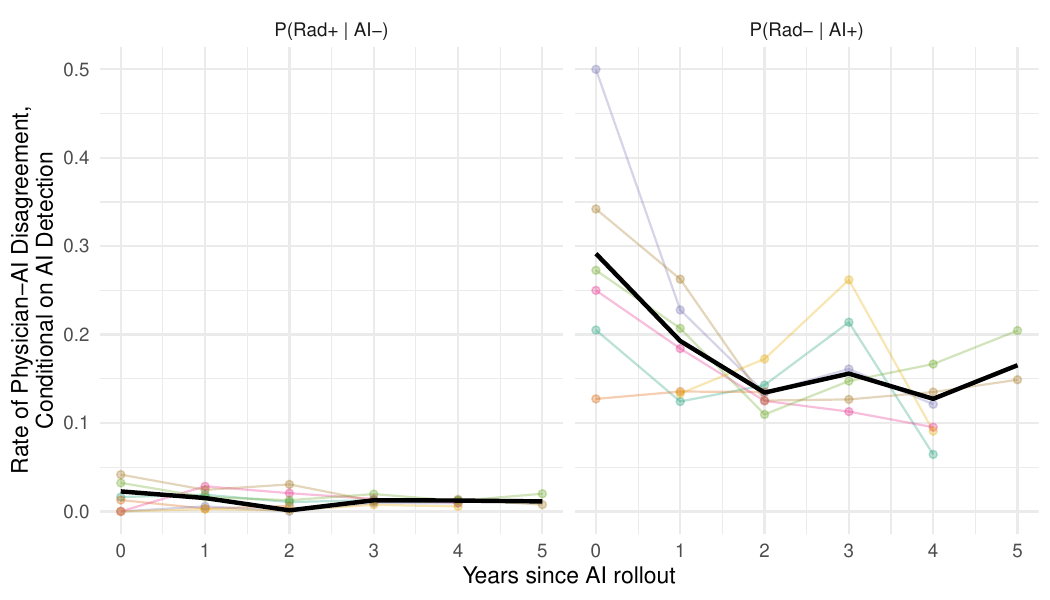}
    \caption{Emergency Department Disagreement Rates Between AI and Radiologists Over Time}
    \label{fig:disagreement_time_ed}
    \medskip
    \begin{minipage}{\textwidth}
    \setlength{\parindent}{0pt}
    This figure shows the rate of disagreement between AI predictions and radiologist diagnoses based on only Emergency Department CTPA scans in the sample. The left panel shows disagreement when the AI detects no PE but radiologists observe one. The right panel shows disagreement when the AI detects PE and the radiologist does not. The colored lines are disagreement patterns by care site per the staggered rollout of the AI system. The solid black lines are simple average per period. Time is measured in years since the initial AI rollout in 2019 per \cref{tab:care_site_summary}. Care site 8 is excluded due to its limited scan volume.
    \end{minipage}
\end{figure}

\clearpage
\begin{figure}[!t]	
    \centering
    \includegraphics[width=.9\textwidth]{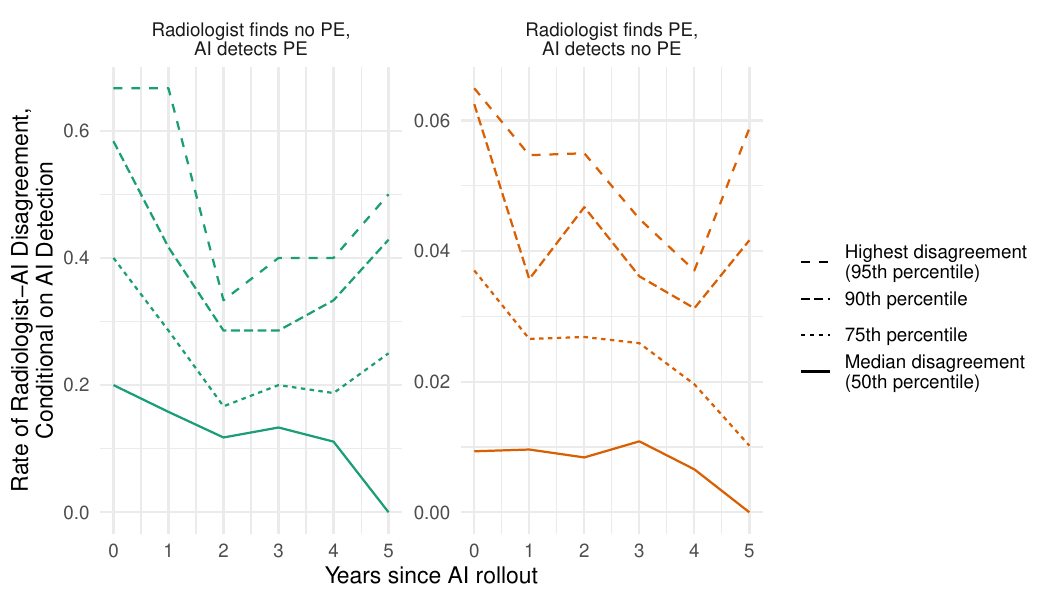}
    \caption{Cross-Sectional Distribution of Radiologist-AI Disagreement Rates Over Time}
    \label{fig:disagreement_distribution}
    \medskip
    \begin{minipage}{\textwidth}
    \setlength{\parindent}{0pt}
    This figure displays the evolution of the cross-sectional distribution of radiologist-AI disagreement rates over five years following the AI deployment. The left panel shows disagreement rates when AI predicts PE, while the right panel shows rates when AI does not predict PE. Lines represent percentiles of the disagreement distribution (50th, 75th, 90th, and 95th) in each year after rollout across all care sites.
    \end{minipage}
\end{figure}
\clearpage

\begin{figure}
    \centering
    \begin{tikzpicture}
\begin{axis}[
    xbar,
    bar width=0.6cm,
    width=14cm,
    height=9cm,
    enlarge y limits=0.15,
    xlabel={\textbf{P(Rad$-$ $|$ AI$+$) (\%)}},
    xlabel style={font=\large},
    ylabel={},
    symbolic y coords={Overall,Q4,Q3,Q2,Q1},
    ytick=data,
    yticklabels={\textbf{Overall}, Q4 (Highest), Q3, Q2, Q1 (Lowest)},
    yticklabel style={font=\normalsize, align=right},
    xtick={0,5,10,15,20,25},
    xmin=0,
    xmax=25,
    xmajorgrids=true,
    grid style={gridcolor, dashed},
    tick label style={font=\normalsize},
    nodes near coords,
    nodes near coords style={font=\small\bfseries, right, xshift=2pt},
    every node near coord/.append style={/pgf/number format/.cd, fixed, precision=1},
    title={\Large\textbf{Radiologist Disagreement with AI-Positive Predictions}},
    title style={yshift=5pt},
    axis lines*=left,
    clip=false,
    legend style={at={(0.97,0.03)}, anchor=south east, font=\small},
]

\addplot[
    fill=barblue,
    draw=barblue!80!black,
    line width=0.5pt,
] coordinates {
    (0,Overall)
    (19.2,Q4)
    (9.1,Q3)
    (12.3,Q2)
    (19.1,Q1)
};

\addplot[
    fill=overallcolor,
    draw=overallcolor!80!black,
    line width=0.5pt,
] coordinates {
    (15.1,Overall)
};

\legend{Quartile, Overall}

\end{axis}


\end{tikzpicture}
    \caption{Radiologist-AI (Dis)agreement by AI System Engagement}
    \label{fig:disagreement_engagement}
    \medskip
    \begin{minipage}{\textwidth}
    \setlength{\parindent}{0pt}
    The figure displays the conditional disagreement rate, \textit{P(Rad$-$ $|$ AI+)}, by quartile of radiologists' AI system engagement based on \cref{tab:disagreement_engagement}.
    \textit{P(Rad$-$ $|$ AI+)} is the probability the radiologist diagnoses no PE given the AI predicted PE.
    \end{minipage}
\end{figure}

\clearpage
\begin{figure}[!p]
    \centering
    
    \begin{subfigure}[b]{0.65\textwidth}
        \centering
        \includegraphics[width=\textwidth]{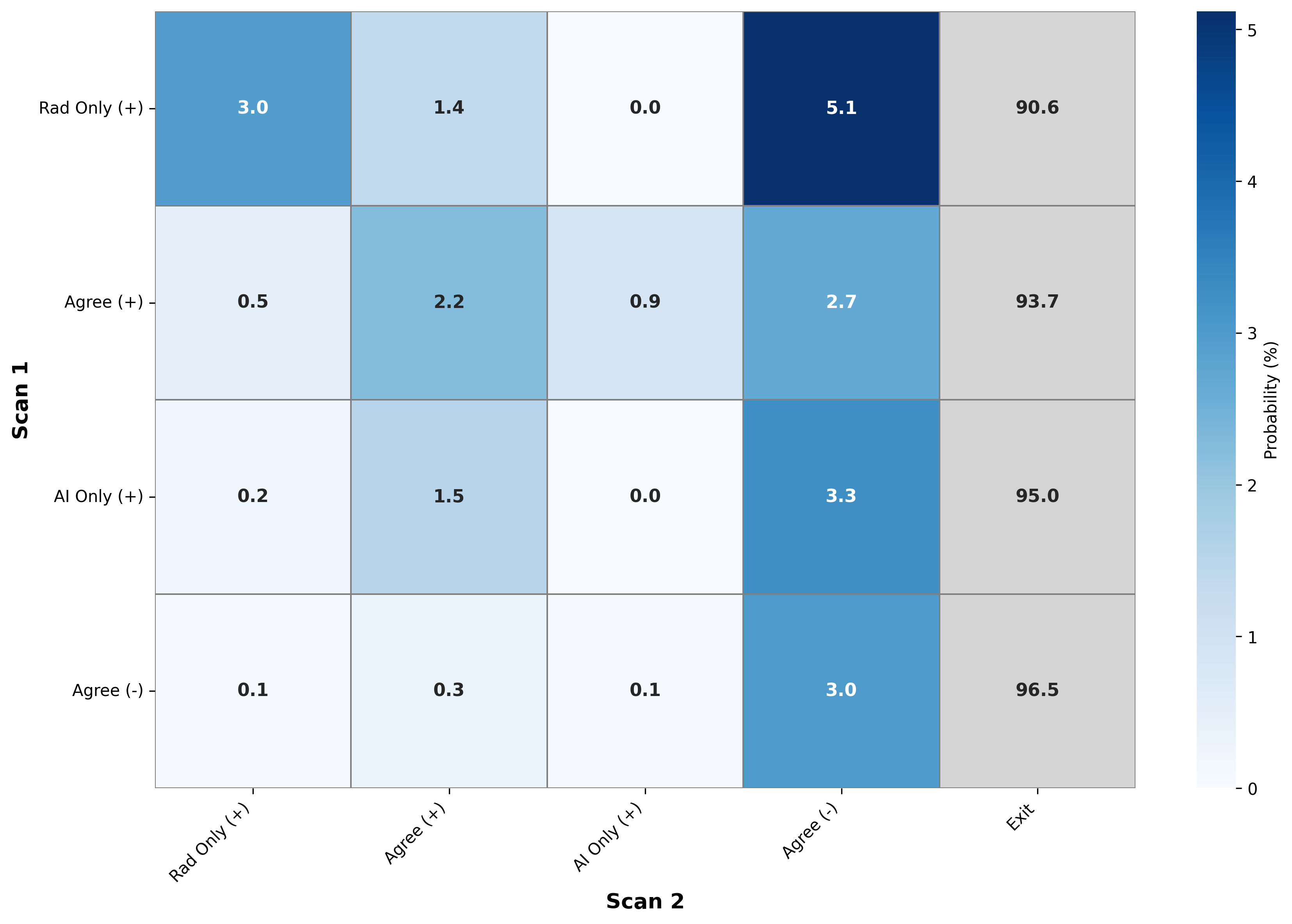}
        \caption{Unconditional Transition Probabilities (Including Exit)}
        \label{fig:transition_heatmap_unconditional}
    \end{subfigure}
    
    \vspace{0.2cm}
    
    \begin{subfigure}[b]{0.65\textwidth}
        \centering
        \includegraphics[width=\textwidth]{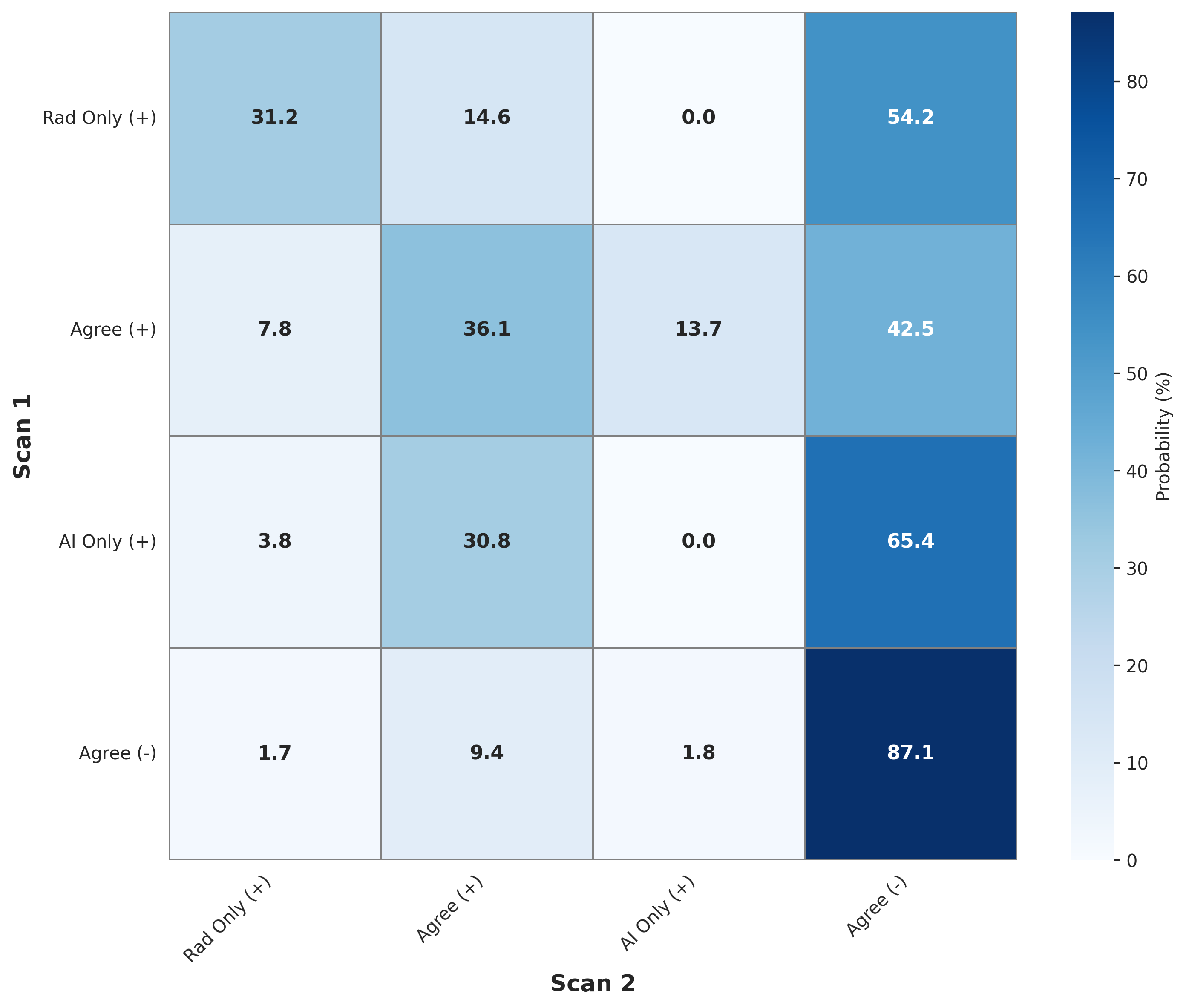}
        \caption{Conditional Transition Probabilities (Excluding Exit)}
        \label{fig:transition_heatmap_conditional}
    \end{subfigure}
    
    \caption{Diagnostic State Transition Probabilities Between Sequential CTPA Scans}
    \label{fig:transition_heatmaps}
    
    \vspace{0.2cm}
    
    \begin{minipage}{\textwidth}
    \setlength{\parindent}{0pt}
    \footnotesize
    The heatmaps show transition probabilities from diagnostic states at the first CTPA scan (Scan 1, rows) to states at the second scan within 30 days (Scan 2, columns). Each cell displays the percentage of patients transitioning from one state to another, with blue shading indicating higher probabilities. Diagnostic states are: Rad Only (+) where only the signing radiologist diagnosed PE; Agree (+) where both the AI and radiologist diagnosed PE; AI Only (+) where only the AI detected PE; and Agree (-) where both found no PE. \Cref{fig:transition_heatmap_unconditional} presents unconditional transition probabilities for all patients with a first scan (N = 84,640), including the Exit column (gray shading) representing patients without a second scan within 30 days. \Cref{fig:transition_heatmap_conditional} shows conditional transition probabilities among only those patients who returned for a second scan within 30 days (N = 4,095), excluding the Exit state.
    \end{minipage}
\end{figure}


\clearpage
\begin{table}[!ht]
    \caption{Care Site Summary Statistics}
    \label{tab:care_site_summary}
    \centering
    \footnotesize
    \renewcommand{\arraystretch}{.85}
    \resizebox{0.95\linewidth}{!}{%
        \begin{tabular}{cccccc}
    \toprule\toprule
    \textbf{Care Site ID} & \textbf{AI Rollout Date} & \textbf{N CTPAs} & \textbf{N Patients} & \textbf{N Radiologists} & \textbf{PE Positivity Rate (\%)} \\
    \midrule
    1 & 08/29/2019 & 47,175 & 35,158 & 252 & 11.3 \\
    2 & 09/10/2019 & 20,556 & 15,768 & 159 & 9.6 \\
    3 & 08/29/2019 & 19,849 & 15,269 & 195 & 11.0 \\
    4 & 11/19/2020 & 11,725 & 9,326 & 20 & 8.3 \\
    5 & 07/12/2022 & 10,305 & 8,535 & 19 & 8.5 \\
    6 & 11/23/2020 & 4,628 & 3,644 & 21 & 6.9 \\
    7 & 07/15/2020 & 2,625 & 2,357 & 85 & 9.6 \\
    8 & 07/14/2021 & 200 & 187 & 26 & 8.5 \\
    \midrule
    \textbf{Total} & — & \textbf{117,063} & \textbf{84,640} & \textbf{388} & \textbf{10.2} \\
    \bottomrule\bottomrule
\end{tabular}

    }
    
    \vspace{0.2cm}
    \begin{minipage}{\textwidth}
        \setlength{\parindent}{0pt}
        The table presents summary statistics for the eight hospital care sites that received the AI system deployment during the sample. Two additional sites within the hospital system were excluded from the sample due to low volume (one scan each during the study period); neither site implemented the AI system. Sites are anonymized and sorted by number of CTPA scans (descending). \textit{AI Rollout Date} indicates when the AI clinical decision support system was first deployed at each site. \textit{N CTPAs} is the total number of CT pulmonary angiogram scans performed at the site during the study period. \textit{N Patients} is the number of unique patients who received CTPAs at the site. \textit{N Radiologists} is the number of unique signing radiologists who interpreted scans at the site. \textit{PE Positivity Rate} is the percentage of scans with a positive pulmonary embolism diagnosis by the signing radiologist across the full sample period. The Total row reports aggregate statistics across all eight AI-receiving sites. \textit{N Patients} and \textit{N Radiologists} in the Total row reflect unique individuals across all sites (not the sum of site-level counts) to avoid double-counting patients and radiologists who receive care or work at multiple sites.
    \end{minipage}
\end{table}

\clearpage
\begin{table}[!ht]
    \caption{Patient Characteristics}
    \label{tab:patient_characteristics}
    \centering
    \footnotesize
    \renewcommand{\arraystretch}{.85}
    \resizebox{0.8\linewidth}{!}{%
        \begin{tabular}{l*{4}{c}}
\toprule\toprule
& Full Sample & Pre-AI Period & AI Period & Diff \\
& $\hat{\mu}_1$ & $\hat{\mu}_2$ & $\hat{\mu}_3$ & $\hat{\mu}_3-\hat{\mu}_2$ \\
& $\hat{\sigma}_1$ & $\hat{\sigma}_2$ & $\hat{\sigma}_3$ & (se) \\
\midrule
\multicolumn{5}{l}{\textit{Panel A: Demographics}} \\[0.5ex]
N patients & 84,640 & 42,612 & 42,028 & -584 \\
&  &  &  &  \\[1ex]
Age & 59.84 & 59.49 & 60.20 & 0.72 \\
& 18.67 & 18.46 & 18.88 & (0.13) \\[1ex]
\% Female & 59.72 & 61.06 & 58.37 & -2.69 \\
& 49.05 & 48.76 & 49.29 & (0.34) \\[1ex]
Race/Ethnicity (\%) &  &  &  &  \\
&  &  &  &  \\[0.3ex]
\quad White & 68.62 & 69.41 & 67.82 & -1.59 \\
& 46.41 & 46.08 & 46.72 & (0.32) \\[1ex]
\quad Black & 17.63 & 17.48 & 17.79 & 0.31 \\
& 38.11 & 37.98 & 38.25 & (0.26) \\[1ex]
\quad Asian & 1.47 & 1.23 & 1.70 & 0.46 \\
& 12.01 & 11.04 & 12.92 & (0.08) \\[1ex]
\quad Hispanic & 13.65 & 12.83 & 14.49 & 1.65 \\
& 34.34 & 33.45 & 35.20 & (0.24) \\[1ex]
\midrule
\multicolumn{5}{l}{\textit{Panel B: Scan Utilization}} \\[0.5ex]
Number of scans & 1.38 & 1.51 & 1.26 & -0.25 \\
& 1.03 & 1.26 & 0.71 & (0.01) \\[1ex]
\% Multiple scans & 22.36 & 27.13 & 17.52 & -9.60 \\
& 41.66 & 44.46 & 38.02 & (0.28) \\[1ex]
\% Scans within 7 days & 1.64 & 1.52 & 1.76 & 0.24 \\
& 12.70 & 12.24 & 13.14 & (0.09) \\[1ex]
\% Scans within 30 days & 5.13 & 4.98 & 5.28 & 0.30 \\
& 22.06 & 21.76 & 22.36 & (0.15) \\[1ex]
\midrule
\multicolumn{5}{l}{\textit{Panel C: PE Diagnosis}} \\[0.5ex]
\% No Rad diagnosed PE on any scan & 88.08 & 87.96 & 88.20 & 0.24 \\
& 32.40 & 32.54 & 32.26 & (0.22) \\[1ex]
\% Rad diagnosed PE on $\geq$1 scan & 11.92 & 12.04 & 11.80 & -0.24 \\
& 32.40 & 32.54 & 32.26 & (0.22) \\[1ex]
\% Rad diagnosed PE on all scans & 7.19 & 6.35 & 8.04 & 1.69 \\
& 25.83 & 24.38 & 27.19 & (0.18) \\[1ex]
\midrule
\multicolumn{5}{l}{\textit{Panel D: Mortality}} \\[0.5ex]
30-day mortality rate & 4.73 & 4.34 & 5.13 & 0.79 \\
& 21.23 & 20.37 & 22.06 & (0.15) \\[1ex]
90-day mortality rate & 7.39 & 7.05 & 7.73 & 0.67 \\
& 26.16 & 25.61 & 26.70 & (0.18) \\[1ex]
1-year mortality rate & 11.77 & 11.71 & 11.84 & 0.14 \\
& 32.23 & 32.15 & 32.31 & (0.22) \\[1ex]
\bottomrule\bottomrule
\end{tabular}
    }
    
    \vspace{0.2cm}
    \begin{minipage}{\linewidth}
        \setlength{\parindent}{0pt}
        The table presents summary statistics for patients who received CTPA scans during the sample period. The Pre-AI period is defined as scans before the AI rollout at each care site; The AI period is defined as scans after the AI rollout at each care site. For each variable, the first row shows sample means ($\hat{\mu}_1$, $\hat{\mu}_2$, and $\hat{\mu}_3$) and their difference ($\hat{\mu}_3 - \hat{\mu}_2$). The second row shows sample standard deviations ($\hat{\sigma}_1$, $\hat{\sigma}_2$, and $\hat{\sigma}_3$) and the heteroskedasticity-robust standard error (se) of the difference between the Pre-AI and AI period. \textit{N patients} shows the number of observations in each sample. All percentages are calculated as shares of total patients in each period. Mortality outcomes are measured from a patient's first CTPA scan date. 
    \end{minipage}
\end{table}

\clearpage
\begin{table}[!ht]
    \caption{CTPA Scan Characteristics}
    \label{tab:scan_characteristics}
    \centering
    \footnotesize
    \renewcommand{\arraystretch}{.85}
    \resizebox{0.75\linewidth}{!}{%
        \begin{tabular}{l*{4}{c}}
\toprule\toprule
& Full Sample & Pre-AI Period & AI Period & Diff \\
& $\hat{\mu}_1$ & $\hat{\mu}_2$ & $\hat{\mu}_3$ & $\hat{\mu}_3-\hat{\mu}_2$ \\
& $\hat{\sigma}_1$ & $\hat{\sigma}_2$ & $\hat{\sigma}_3$ & (se) \\
\midrule
\multicolumn{5}{l}{\textit{Panel A: Scan Context}} \\[0.5ex]
N scans & 117,065 & 54,150 & 62,915 & 8,765 \\
&  &  &  &  \\[1ex]
\% Repeat scan within 7 days & 1.25 & 0.96 & 1.51 & 0.54 \\
& 3.25 & 4.19 & 4.85 & (0.06) \\[1ex]
\% Repeat scan within 30 days & 4.49 & 3.59 & 5.27 & 1.68 \\
& 6.06 & 8.00 & 8.91 & (0.12) \\[1ex]
\% Emergency Department & 66.00 & 56.10 & 74.52 & 18.43 \\
& 13.85 & 21.33 & 17.37 & (0.28) \\[1ex]
\midrule
\multicolumn{5}{l}{\textit{Panel B: Diagnostic Outcomes}} \\[0.5ex]
\% Radiologist PE (+) & 10.19 & 10.11 & 10.25 & 0.14 \\
& 8.84 & 12.96 & 12.09 & (0.18) \\[1ex]
\% AI PE (+) & -- & -- & 10.66 & -- \\
& -- & -- & 13.29 & -- \\[1ex]
\midrule
\multicolumn{5}{l}{\textit{Panel C: Radiologist-AI (Dis)agreement}} \\[0.5ex]
\% AI Only (+) & -- & -- & 1.74 & -- \\
& -- & -- & 5.63 & -- \\[1ex]
\% Rad Only (+) & -- & -- & 1.57 & -- \\
& -- & -- & 5.36 & -- \\[1ex]
\% Agree (+) & -- & -- & 8.92 & -- \\
& -- & -- & 12.28 & -- \\[1ex]
\% Agree (-) & -- & -- & 87.77 & -- \\
& -- & -- & 14.12 & -- \\[1ex]
\midrule
\multicolumn{5}{l}{\textit{Panel D: Workflow}} \\[0.5ex]
\% Day shift (7am-3pm) & 15.46 & 15.23 & 15.66 & 0.43 \\
& 10.57 & 15.44 & 14.49 & (0.21) \\[1ex]
\% Evening shift (3pm-11pm) & 51.02 & 51.56 & 50.55 & -1.01 \\
& 14.61 & 21.48 & 19.93 & (0.29) \\[1ex]
\% Night shift (11pm-7am) & 33.52 & 33.21 & 33.79 & 0.58 \\
& 13.80 & 20.24 & 18.86 & (0.28) \\[1ex]
\bottomrule\bottomrule
\end{tabular}
    }
    
    \vspace{0.2cm}
    \begin{minipage}{\linewidth}
        \setlength{\parindent}{0pt}
        The table presents summary statistics for CTPA scans during the sample period. The Pre-AI period is defined as scans before the AI rollout at each care site; The AI period is defined as scans after the AI rollout at each care site. For each variable, the first row shows sample means ($\hat{\mu}_1$, $\hat{\mu}_2$, and $\hat{\mu}_3$) and their difference ($\hat{\mu}_3 - \hat{\mu}_2$). The second row shows sample standard deviations ($\hat{\sigma}_1$, $\hat{\sigma}_2$, and $\hat{\sigma}_3$) and the heteroskedasticity-robust standard error (se) of the difference. \textit{N scans} shows the number of observations in each sample. \textit{\% AI PE (+)} (Panel B) and all \textit{Diagnostic Discordance} metrics (Panel C) are calculated only on scans with AI predictions available (n=53,880, representing 85.6\% of AI period scans). \textit{AI Only (+)} indicates cases where AI detected PE but radiologist did not; \textit{Rad Only (+)} indicates cases where radiologist detected PE but AI did not; \textit{Agree (+)} and \textit{Agree (-)} indicate agreement on PE presence and absence, respectively.
    \end{minipage}
\end{table}

\clearpage
\begin{table}[!ht]
    \caption{Radiologist Characteristics}
    \label{tab:radiologist_characteristics}
    \centering
    \footnotesize
    \renewcommand{\arraystretch}{.75}
    \resizebox{0.75\linewidth}{!}{%
        \begin{tabular}{l*{3}{c}}
\toprule\toprule
& Full Sample & Pre-AI Period & AI Period \\
\midrule
\multicolumn{4}{l}{\textit{Panel A: Reading Volume}} \\[0.5ex]
N signing radiologists & 389 & 309 & 187 \\[0.5ex]
\textit{Scans read per radiologist per month:} & & & \\
\quad Mean & 6.8 & 5.3 & 10.4 \\
\quad Median & 3.6 & 3.2 & 5.4 \\
\quad 5th percentile & 1.0 & 1.0 & 1.0 \\
\quad 25th percentile & 1.8 & 1.5 & 3.0 \\
\quad 75th percentile & 7.9 & 6.8 & 17.7 \\
\quad 95th percentile & 24.2 & 15.8 & 30.3 \\
\midrule
\multicolumn{4}{l}{\textit{Panel B: Demographics}} \\[0.5ex]
\% Female & 32.11 & 32.62 & 30.38 \\
Years since medical school & 12.8 & 13.3 & 15.8 \\
Years active in sample & 2.2 & 1.7 & 1.7 \\
\midrule
\multicolumn{4}{l}{\textit{Panel C: PE Diagnosis Rate}} \\[0.5ex]
\quad Mean & 11.26 & 10.45 & 12.05 \\
\quad Median & 9.15 & 8.74 & 10.13 \\
\quad 25th percentile & 4.35 & 0.00 & 7.14 \\
\quad 75th percentile & 12.50 & 12.24 & 12.97 \\
\midrule
\multicolumn{4}{l}{\textit{Panel D: AI-Radiologist Interaction}} \\[0.5ex]
\textit{Conditional agreement rates:} & & & \\
\quad P(Rad+ $|$ AI+) (\%) & --- & --- & 84.1 \\
\quad P(Rad$-$ $|$ AI$-$) (\%) & --- & --- & 97.2 \\[1ex]
\textit{Conditional disagreement rates:} & & & \\
\quad P(Rad$-$ $|$ AI+) (\%) & --- & --- & 15.9 \\
\quad P(Rad+ $|$ AI$-$) (\%) & --- & --- & 2.8 \\[1ex]
\bottomrule\bottomrule
\end{tabular}%
    }

    \vspace{0.2cm}
    \begin{minipage}{\linewidth}
        \setlength{\parindent}{0pt}
        The table presents summary statistics across individual signing radiologists over the sample period. The sample includes only the eight care sites that received AI system deployment. The Pre-AI period includes radiologists active before the AI rollout at their respective care sites; the AI period includes radiologists active after the AI rollout. Some radiologists appear in both periods. \textit{Scans read per radiologist per month} is calculated as total scans divided by the number of unique year-month combinations in which each radiologist read CTPA scans. \textit{Mean years active in sample} is calculated as the number of unique year-month combinations in which each radiologist read CTPA scans, divided by 12 to convert to years. \textit{Mean years since medical school} is calculated for each radiologist as the average across all their scans of (scan year - medical school graduation year), providing a measure of average experience level during the period. \textit{Conditional agreement rates} show the average percentage of scans where radiologists agreed with the AI, calculated separately for AI-positive and AI-negative predictions. \textit{P(Rad+ | AI+)} is the probability the radiologist diagnoses PE given the AI predicted PE; \textit{P(Rad- | AI-)} is the probability the radiologist diagnoses no PE given the AI predicted no PE; \textit{P(Rad- | AI+)} is the probability the radiologist diagnoses no PE given the AI predicted PE; \textit{P(Rad+ | AI-)} is the probability the radiologist diagnoses PE given the AI predicted no PE. Demographic information is obtained from CMS National Plan and Provider Enumeration System (NPPES) data matched to signing radiologist identifiers.
    \end{minipage}
\end{table}

\clearpage
\begin{table}[!ht]
    \caption{Diagnostic Efficiency}
    \label{tab:scan_efficiency}
    \centering
    \footnotesize
    \renewcommand{\arraystretch}{.85}
    \resizebox{0.9\linewidth}{!}{%
        \begin{tabular}{l*{4}{c}}
\toprule\toprule
& Full Sample & Pre-AI Period & AI Period & Diff \\
& $\hat{\mu}_1$ & $\hat{\mu}_2$ & $\hat{\mu}_3$ & $\hat{\mu}_3-\hat{\mu}_2$ \\
& $\hat{\sigma}_1$ & $\hat{\sigma}_2$ & $\hat{\sigma}_3$ & (se) \\
\midrule
N scans & 117,065 & 54,150 & 62,915 & 8,765 \\
&  &  &  &  \\[1ex]
Hours from Order to Diagnosis & 3.65 & 3.57 & 3.72 & 0.15 \\
& 12.00 & 12.55 & 11.51 & (0.07) \\[1ex]
Hours from Order to Diagnosis: Rad PE (+) & 3.43 & 3.23 & 3.61 & 0.39 \\
& 10.10 & 9.95 & 10.22 & (0.19) \\[1ex]
Hours from Order to Diagnosis: Rad PE (-) & 3.67 & 3.61 & 3.72 & 0.11 \\
& 12.17 & 12.80 & 11.60 & (0.08) \\[1ex]
Hours from Order to Diagnosis: AI Only (+) & -- & -- & 3.99 & -- \\
& -- & -- & 10.65 & -- \\[1ex]
Hours from Order to Diagnosis: Rad Only (+) & -- & -- & 3.78 & -- \\
& -- & -- & 10.35 & -- \\[1ex]
Hours from Order to Diagnosis: Agree (+) & -- & -- & 3.61 & -- \\
& -- & -- & 10.17 & -- \\[1ex]
Hours from Order to Diagnosis: Agree (-) & -- & -- & 3.74 & -- \\
& -- & -- & 11.59 & -- \\[1ex]
\bottomrule\bottomrule
\end{tabular}%
    }
    
    \vspace{0.2cm}
    \begin{minipage}{\linewidth}
        \setlength{\parindent}{0pt}
        The table presents efficiency metrics for CTPA scans during the sample period. The Pre-AI period is defined as scans before the AI rollout at each care site; The AI period is defined as scans after the AI rollout at each care site. \textit{Hours from Order to Diagnosis} excludes scans with negative duration or that exceed 7 days. These observations are set to missing but retained in the sample for \textit{N scans}. For each variable, the first row shows sample means ($\hat{\mu}_1$, $\hat{\mu}_2$, and $\hat{\mu}_3$) and their difference ($\hat{\mu}_3 - \hat{\mu}_2$). The second row shows sample standard deviations ($\hat{\sigma}_1$, $\hat{\sigma}_2$, and $\hat{\sigma}_3$) and the heteroskedasticity-robust standard error (se) of the difference. \textit{Rad PE (+)} and \textit{Rad PE (-)} indicate radiologist diagnosis of PE presence or absence, respectively. The four discordance categories are calculated only on scans with AI predictions available (n=53,880, representing 85.6\% of AI period scans). \textit{AI Only (+)} indicates cases where AI detected PE but radiologist did not; \textit{Rad Only (+)} indicates cases where radiologist detected PE but AI did not; \textit{Agree (+)} and \textit{Agree (-)} indicate agreement on PE presence and absence, respectively.
    \end{minipage}
\end{table}

\clearpage
\begin{table}[!ht]
    \caption{Radiologist Engagement with AI: Descriptive Statistics}
    \label{tab:engagement_descriptive}
    \centering
    \footnotesize
    \renewcommand{\arraystretch}{1}
    \resizebox{0.5\linewidth}{!}{%
        \begin{tabular}{l*{4}{c}}
\toprule\toprule
& Mean & & Median & \\
& (SD) & & [IQR] & \\
\midrule
N radiologists & 148 & & & \\[1ex]
\midrule
\multicolumn{5}{l}{\textit{Panel A: Volume \& Engagement}} \\[0.5ex]
Total AI notifications & 702 & & 239 & \\
& (1106) & & [31, 946] & \\[1ex]
AI hover interactions & 247 & & 71 & \\
& (475) & & [9, 267] & \\[1ex]
Engagement rate (\%) & 33.9 & & 32.4 & \\
& (19.9) & & [20.7, 45.2] & \\[1ex]
Hover duration (minutes) & 1.00 & & 1.00 & \\
& (0.00) & & [1.00, 1.00] & \\[1ex]
\midrule
\multicolumn{5}{l}{\textit{Panel B: AI Triage Efficiency}} \\[0.5ex]
AI triage lead (minutes) & 20.0 & & 8.8 & \\
& (45.5) & & [6.4, 13.6] & \\[1ex]
Reaction time (minutes) & 5.3 & & 4.0 & \\
& (6.0) & & [2.3, 5.5] & \\[1ex]
\% Cases opened pre-AI alert & 98.2 & & 99.7 & \\
& (2.6) & & [97.4, 100.0] & \\[1ex]
\midrule
\multicolumn{5}{l}{\textit{Panel C: Reading Efficiency}} \\[0.5ex]
Open-to-draft time (minutes) & 2.2 & & 0.0 & \\
& (5.9) & & [0.0, 0.2] & \\[1ex]
Draft-to-final time (minutes) & 18.0 & & 14.7 & \\
& (15.9) & & [7.9, 22.2] & \\[1ex]
Total time-to-final (minutes) & 53.9 & & 9.1 & \\
& (224.2) & & [6.6, 12.2] & \\[1ex]
\midrule
\multicolumn{5}{l}{\textit{Panel D: Engagement \& Efficiency: AI+ Cases}} \\[0.5ex]
Time-to-final: AI+ with hover & 30.1 & & 17.0 & \\
& (85.8) & & [12.1, 22.8] & \\[1ex]
Time-to-final: AI+ no hover & 52.1 & & 18.2 & \\
& (106.2) & & [10.9, 28.9] & \\[1ex]
\quad Engagement time savings & 25.7 & & 1.4 & \\
& (96.4) & & [-3.1, 7.9] & \\[1ex]
\bottomrule\bottomrule
\end{tabular}%
    }
    
    \vspace{0.2cm}
    \begin{minipage}{\linewidth}
        \setlength{\parindent}{0pt}
        The table presents radiologist-level descriptive statistics for AI system engagement during the AI period. The sample is restricted to notifications where the receiving physician signed the final report (using both the report\_signer boolean flag and physician-signer name matching), ensuring analysis focuses on cases where the notified radiologist had primary diagnostic responsibility. Statistics are calculated across 148 radiologists who received and signed AI notifications. For each metric, we report both the mean (with standard deviation) and median (with interquartile range) to characterize central tendency and dispersion in the face of potential outliers. \textit{Panel A} describes overall system usage: total notifications received, hover interactions (where radiologists actively viewed AI predictions), engagement rate (\% of notifications with hover), and typical hover duration. \textit{Panel B} measures AI's triage function: the time AI alerted radiologists before they would have independently opened cases (triage lead), radiologists' response time after receiving alerts (reaction time), and how often radiologists had already opened cases before AI alerts arrived (preemptive opens). \textit{Panel C} captures reading efficiency through the workflow stages: time from opening a case to starting the draft, time from draft to final sign-off, and total time from alert to finalization. \textit{Panel D} examines whether AI engagement affects workflow for AI-positive cases by comparing time-to-finalization when radiologists did versus did not hover over AI predictions; the engagement time savings represents the within-radiologist difference (no-hover minus hover time). 
    \end{minipage}
\end{table}

\clearpage
\begin{table}[!ht]
    \caption{Radiologist-AI (Dis)agreement by Scan Timing}
    \label{tab:disagreement_shift}
    \centering
    \footnotesize
    \renewcommand{\arraystretch}{.85}
    \resizebox{0.95\linewidth}{!}{%
        \begin{tabular}{l*{6}{c}}
\toprule\toprule
& \multicolumn{6}{c}{(Dis)agreement Pattern (\%)} \\
\cmidrule(lr){2-7}
& & & \multicolumn{4}{c}{Conditional Rates} \\
\cmidrule(lr){4-7}
Shift & N Scans & \% ER & P(Rad+$|$AI+) & P(Rad$-$$|$AI$-$) & P(Rad$-$$|$AI+) & P(Rad+$|$AI$-$) \\
\midrule
\multicolumn{7}{l}{\textit{Panel A: By Time of Day}} \\[0.5ex]
Day & 8,395 & 76.5 & 81.5 & 98.2 & 18.5 & 1.8 \\
Evening & 27,193 & 67.6 & 83.7 & 98.2 & 16.3 & 1.8 \\
Night & 18,292 & 83.2 & 84.6 & 98.3 & 15.4 & 1.7 \\
\midrule
\multicolumn{7}{l}{\textit{Panel B: By Day of Week}} \\[0.5ex]
Weekday & 40,577 & 72.9 & 83.9 & 98.3 & 16.1 & 1.7 \\
Weekend & 13,303 & 78.3 & 83.0 & 98.2 & 17.0 & 1.8 \\
\midrule
\textit{Overall} & 53,880 & 74.3 & 83.7 & 98.2 & 16.3 & 1.8 \\
\bottomrule\bottomrule
\end{tabular}
    }  
    
    \vspace{0.2cm}
    \begin{minipage}{\linewidth}
        \setlength{\parindent}{0pt}
        The table presents scan-level radiologist-AI agreement patterns by scan timing (shift) for 53,880 CTPA scans in the AI period with both AI and radiologist diagnoses available. Shifts are defined as Day (7am-3pm), Evening (3pm-11pm), and Night (11pm-7am). Weekend includes Saturday and Sunday. \textit{P(Rad+ $|$ AI+)} is the probability the radiologist diagnoses PE given the AI predicted PE; \textit{P(Rad$-$ $|$ AI$-$)} is the probability the radiologist diagnoses no PE given the AI predicted no PE; \textit{P(Rad$-$ $|$ AI+)} is the probability the radiologist diagnoses no PE given the AI predicted PE; \textit{P(Rad+ $|$ AI$-$)} is the probability the radiologist diagnoses PE given the AI predicted no PE. \textit{\% ER} shows the percentage of scans from the Emergency Department. 
    \end{minipage}
\end{table}

\clearpage
\begin{table}[!ht]
    \caption{Radiologist-AI (Dis)agreement by Demographics}
    \label{tab:disagreement_demographics}
    \centering
    \footnotesize
    \renewcommand{\arraystretch}{.85}
    \resizebox{0.95\linewidth}{!}{%
        \begin{tabular}{l*{6}{c}}
\toprule\toprule
& \multicolumn{6}{c}{(Dis)agreement Pattern (\%)} \\
\cmidrule(lr){2-7}
& & & \multicolumn{4}{c}{Conditional Rates} \\
\cmidrule(lr){4-7}
Group & N Rads & N Scans & P(Rad+$|$AI+) & P(Rad$-$$|$AI$-$) & P(Rad$-$$|$AI+) & P(Rad+$|$AI$-$) \\
\midrule
\multicolumn{7}{l}{\textit{Panel A: By Gender}} \\[0.5ex]
Male & 101 & 35,621 & 82.0 & 97.8 & 18.0 & 2.2 \\
Female & 45 & 14,268 & 88.4 & 96.0 & 11.6 & 4.0 \\
\midrule
\multicolumn{7}{l}{\textit{Panel B: By Experience (Years Since Medical School)}} \\[0.5ex]
Q1 (0-5y) & 39 & 2,069 & 83.4 & 95.6 & 16.6 & 4.4 \\
Q2 (6-10y) & 37 & 14,654 & 87.2 & 96.8 & 12.8 & 3.2 \\
Q3 (11-20y) & 33 & 17,662 & 85.1 & 98.6 & 14.9 & 1.4 \\
Q4 (21+y) & 37 & 15,504 & 80.6 & 98.1 & 19.4 & 1.9 \\
\midrule
\multicolumn{7}{l}{\textit{Panel C: By Scan Volume}} \\[0.5ex]
Q1 (Low) & 44 & 300 & 84.1 & 94.7 & 15.9 & 5.3 \\
Q2 & 41 & 1,620 & 85.3 & 97.6 & 14.7 & 2.4 \\
Q3 & 42 & 10,719 & 83.4 & 98.4 & 16.6 & 1.6 \\
Q4 (High) & 42 & 41,241 & 83.7 & 98.2 & 16.3 & 1.8 \\
\midrule
\textit{Overall} & 169 & 53,880 & 84.1 & 97.2 & 15.9 & 2.8 \\
\bottomrule\bottomrule
\end{tabular}%
    }
    
    \vspace{0.2cm}
    \begin{minipage}{\linewidth}
        \setlength{\parindent}{0pt}
        The table presents radiologist-AI agreement patterns by demographic characteristics across 169 individual signing radiologists in the AI period. Statistics are calculated at the radiologist level: for each radiologist, conditional agreement rates are calculated based on their individual scans, then averaged across radiologists within each demographic group, treating each radiologist equally regardless of their scan volume. Gender information is obtained from CMS National Plan and Provider Enumeration System (NPPES) data. Experience is measured as years since medical school graduation at the time of each scan. \textit{P(Rad+ $|$ AI+)} is the probability the radiologist diagnoses PE given the AI predicted PE; \textit{P(Rad$-$ $|$ AI$-$)} is the probability the radiologist diagnoses no PE given the AI predicted no PE; \textit{P(Rad$-$ $|$ AI+)} is the probability the radiologist diagnoses no PE given the AI predicted PE; \textit{P(Rad+ $|$ AI$-$)} is the probability the radiologist diagnoses PE given the AI predicted no PE. 
    \end{minipage}
    \end{table}

\clearpage
\begin{table}[!ht]
    \caption{Radiologist-AI (Dis)agreement by AI System Engagement}
    \label{tab:disagreement_engagement}
    \centering
    \footnotesize
    \renewcommand{\arraystretch}{.85}
    \resizebox{0.95\linewidth}{!}{%
        \begin{tabular}{l*{6}{c}}
\toprule\toprule
& \multicolumn{6}{c}{(Dis)agreement Pattern (\%)} \\
\cmidrule(lr){2-7}
& & & \multicolumn{4}{c}{Conditional Rates} \\
\cmidrule(lr){4-7}
Quartile & N Rads & Avg. Engagement & P(Rad+$|$AI+) & P(Rad$-$$|$AI$-$) & P(Rad$-$$|$AI+) & P(Rad+$|$AI$-$) \\
& & Rate (\%) & & & & \\
\midrule
Q1 (Lowest) & 39 & 8.3 & 80.9 & 97.2 & 19.1 & 2.8 \\
Q2 & 38 & 17.8 & 87.7 & 96.0 & 12.3 & 4.0 \\
Q3 & 38 & 26.3 & 90.9 & 98.6 & 9.1 & 1.4 \\
Q4 (Highest) & 38 & 44.5 & 80.8 & 96.5 & 19.2 & 3.5 \\
\midrule
\textit{Overall} & 153 & 24.1 & 84.9 & 97.1 & 15.1 & 2.9 \\
\bottomrule\bottomrule
\end{tabular}
    }
    
    \vspace{0.2cm}
    \begin{minipage}{\linewidth}
        \setlength{\parindent}{0pt}
        The table presents radiologist-AI (dis)agreement patterns by quartile of average engagement rate across 153 individual signing radiologists in the AI period. Engagement rate is calculated as the proportion of positive PE notifications from the AI that the radiologist hovered over before signing the report, averaged across all scans for each radiologist. Statistics are calculated at the radiologist level: for each radiologist, conditional agreement rates are calculated based on their individual scans, then averaged across radiologists within each engagement quartile, treating each radiologist equally regardless of their scan volume. \textit{P(Rad+ $|$ AI+)} is the probability the radiologist diagnoses PE given the AI predicted PE; \textit{P(Rad$-$ $|$ AI$-$)} is the probability the radiologist diagnoses no PE given the AI predicted no PE; \textit{P(Rad$-$ $|$ AI+)} is the probability the radiologist diagnoses no PE given the AI predicted PE; \textit{P(Rad+ $|$ AI$-$)} is the probability the radiologist diagnoses PE given the AI predicted no PE.
    \end{minipage}
\end{table}

\end{document}